\documentclass{article}

\usepackage{graphicx}
\usepackage{caption}
\usepackage{float}
\usepackage{subfigure}
\usepackage{epstopdf}
\usepackage{amsmath,amssymb}
\usepackage{verbatim}
\usepackage{multirow}
\usepackage{comment}
\usepackage{color}

\begin{document}
\title{Elastic scattering of protons from $\sqrt{s}=23.5$ GeV to 7 TeV from a generalized Bialas-Bzdak model\footnote{Dedicated to the 70th birthday of Karsten Eggert,  outgoing Spokesman of TOTEM}}

\author{
T. Cs\"org\H{o}\footnote{csorgo.tamas@wigner.mta.hu}\\
Wigner Research Centre for Physics, Hungarian Academy of Sciences \\ H-1525 Budapest 114, P.O.Box 49, Hungary
\vspace{8mm}\\
F. Nemes\footnote{frigyes.janos.nemes@cern.ch}\\
CERN, CH-1211 Geneva 23, Switzerland
}

\maketitle

\begin{abstract}
The Bialas-Bzdak model of elastic proton-proton scattering is generalized 
to the case when the real part of the parton-parton level forward 
scattering amplitude is non-vanishing. Such a generalization enables the model
to describe well the dip region of the differential cross-section of 
elastic scattering at the ISR energies, and improves significantly the 
ability of the model to describe also the recent TOTEM data 
at $\sqrt{s} = 7$ TeV LHC energy. 
Within this framework, both the increase of the total cross-section, 
as well as the decrease of the location of the dip with increasing colliding 
energies, is related to the increase of the quark-diquark distance and 
to the increase of the ``fragility" of the protons with increasing energies.
In addition, we present and test
the validity of  two new phenomenological relations: 
one of them relates the total p+p cross-section to an effective,
model-independent proton radius, while the other relates the position 
of the dip in the differential elastic cross-section to the measured value of
the total cross-section. 
\end{abstract}

\newpage

\section{Introduction}
Diffractive scattering of energetic electrons on various nuclei
allowed Hofstadter and collaborators to determine the radius and the surface thickness of electric charge distribution
inside the nuclei, and also resulted in two
simple phenomenological formulae: namely, that the charge radius scales with 
increasing mass number as (1.07 $\pm$ 0.02)$\times A^{1/3}$ fm, 
while the surface thickness, within the errors of the observation, is a constant of
(2.4 $\pm$ 0.3) fm~\cite{Hahn:1956zz}.
By increasing the momentum of the elastically scattered particle one can probe
deeper and deeper details of matter, due to the de Broglie duality $\lambda  = \hbar / p$
between particle properties like momentum $p$ and wave-like properties 
like wavelength $\lambda $. To resolve the internal charge distribution inside atomic nuclei, Hofstadter and
colleagues used electrons with bombarding energies of $ E = 153 - 183 $ MeV, corresponding to spatial resolutions of
 $\lambda \approx 1-2$ fm. Currently, elastic scattering of protons on protons has been experimentally investigated
by the CERN LHC experiment TOTEM, at the colliding energies of $\sqrt{s} = 7$ TeV, corresponding to a spatial resolution in
 the center of mass system of $\lambda \approx 0.05$ fm, which is sufficiently small to  study the internal structure of protons.
 
In the present paper  we exploit these experimental results and check how far can one
progress with the qualitative as well as quantitative description of the differential cross section of elastic proton-proton scattering at ISR as well as at CERN LHC
colliding energies, covering an impressive range from $\sqrt{s} = 23.5$ GeV to 7 TeV.
Our investigations were motivated not only by the first experimental results of TOTEM 
on elastic proton-proton collisions~\cite{Antchev:2011zz,Antchev:2011vs}, but also by an inspiring series of theoretical papers by A. Bialas and
A. Bzdak~\cite{Bialas:2006qf,Bzdak:2007qq,Bialas:2006kw,Bialas:2007eg}, 
that considered pion-proton, proton-proton and nucleus-nucleus scattering using a geometrical picture. 

In the present investigation, following and improving the model of Bialas and 
Bzdak~\cite{Bialas:2006qf,Bzdak:2007qq,Bialas:2006kw,Bialas:2007eg} a quark-diquark based geometrical model is 
used to describe the ISR and the recent LHC TOTEM data. 
We will refer to this Bialas Bzdak model as the BB model or BB, for convenience.

Note also that a  diquark or correlated quark-quark structure has to be considered, given that a similar model with three 
independent, uncorrelated constituent quarks was not able to describe the elastic pp scattering data, as pointed out
already in Ref.~\cite{Bialas:1977xp}. 

By definition, the earlier considered BB model has no real part, and consequently
it is singular at the diffractive minimum. The diffractive minimum, the \textit{dip}, was first seen in the data
collected with the CERN Intersecting Storage Rings (ISR) \cite{Nagy:1978iw,Amaldi:1979kd}. The TOTEM data shows that the dip is shifted 
towards smaller $\left|t\right|$ values at $\sqrt{s}=7$ TeV, consequently with smaller experimental errors, and these data
also exclude a possible vanishing value of the differential cross-section at this dip position.

Therefore to describe the data, including the dip, is a challenging task. We were able to fit the  singular BB model
to the ISR data only in the case, when 3 data points were left out from the data at the dip, and even this did not help at
LHC energies: the BB model simply failed in the dip region of  TOTEM data~\cite{Nemes:2012cp}. 
A recent study~\cite{Martin:2011gi}, which extended the parton approach, using gluon cascades, from hard processes to 
describe high-energy soft and semihard processes, was also  not able to describe the TOTEM data.
Its 3-channel eikonal extension with only one Pomeron missed a quantitative data description
at LHC energies~\cite{Ryskin:2012az} just as well. 
However, as it was demonstrated recently in Ref.~\cite{Khoze:2013dha}
using the Good-Walker formalism with a single, effective Pomeron, it is possible to describe the
pp elastic differential cross-section and the energy dependence of the total cross-section $\sigma_{tot}$,
as well as proton dissociation into low mass system in a large energy range from CERN-ISR to LHC energies.
 
In another recent study, that we become aware during the finalization phase of our manuscript,
TOTEM data on the differential cross-section of elastic pp scattering at $\sqrt{s}= 7 $ TeV were described by a relatively simple
parametrization~\cite{Fagundes:2013aja}
that includes two exponentials and a phase, where the first exponential is suitably modified to take into account the
detailed structure of forward scattering and to include the description of the electromagnetic form-factor of the protons.

A multi-channel eikonal model was considered and found to describe well TOTEM elastic scattering data $d\sigma/dt$ at 7 TeV and also
pp scattering at 53 GeV ISR energies with some interesting physics conclusions: in particular, Ref.~\cite{Lipari:2013kta}
pointed out that elastic, diffractive and total pp scattering cross-sections at 7 TeV within errors saturate the so-called
Miettinen-Pumplin bound~\cite{Miettinen:1978ab}, namely that
$\frac{(\sigma_{el} + \sigma_{diff})}{\sigma_{tot} } \le \frac{1}{2}$.
This result is consistent with very large parton-parton scattering cross-sections at these energies.

A detailed analysis of the real and the imaginary part of the forward scattering amplitude was presented recently in 
Ref.~\cite{Kohara:2013bua,Ferreira:2012zi}, where a regular dependence of the fit parameters as a function of $\ln(s)$ was observed, and this behaviour was found to be
nearly continuous even at the lower, ISR energies, but pointing out the opening of a new mechanism in the energy range of 
$100 < \sqrt{s} < 500$ GeV as indicated by a perturbation of the otherwise smooth energy dependence of the model parameters.
A qualitatively similar observation was also made in Ref.~\cite{Wibig:2011iw}, that noted that the geometrical picture of pp collisions
should be modified considerably when entering the domain of ultra-high energy collisions, and pointed out that TOTEM data at
$\sqrt{s} = 7 $ TeV already indicate the onset of a transition to this region.

In our present study, we investigate in detail a geometrical model of proton-proton collisions, and indeed
find that a qualitative change in elastic pp scattering can be observed when moving from the 
top ISR energy of 63 GeV to the LHC energy of 7 TeV, which is particularly important when the study is focusing
on the dip region. We perform our analysis using a generalized Bialas-Bzdak model.
The BB model originates in Glauber models, which were successfully developed to provide a geometrical as well as 
quantum-optical picture of nucleus-nucleus collisions~\cite{Glauber:1955qq}. The 
cross-section is calculated by taking into account all possible combinations of individual inelastic 
processes~\cite{Glauber:2006gd,Glauber}. In the present work the BB model is supplemented with real part
following Ref.~\cite{Glauber:1984su,Glauber:1987sf}. Therefore the ratio of the real to imaginary part of 
the forward scattering amplitude at zero momentum transfer, $\rho$ is not zero. We refer to our model
version as $\alpha BB$. In total four free model parameters are used in $\alpha BB$, which is rather small. The QCD-based, well-known model of 
Ref.~\cite{Donnachie:2011aa}, where the TOTEM data is refitted successfully with the help of hard Pomeron exchange, 
uses 15 free parameters. Another relevant QCD inspired model applies 9 free parameters, where the Pomeron 
intercept and the interesting dynamical gluon mass are the most important~\cite{Fagundes:2011wn,Fagundes:2011zx}. 

We present two, intuitively found phenomenological relations also. The first one suggests that the total cross-section
can be estimated using an effective proton radius, which is a combination of our three radius fit parameters. We
tested this formula with the singular, original version of the  BB model earlier~\cite{Nemes:2012cp} and now we present its test
with the non-singular, generalized  $\alpha BB$ model. Similarly to the relation of the nuclear radius to the mass number found by Hofstadter,
this relation is found to be only approximate, but giving a good estimate of the total cross section and also its increase with increasing
energies can be related to the modification of the internal structure of protons with increasing colliding energies.
The second relation shows that the position of the dip, $t_{dip}$, is related to the total cross-section, so given 
the total pp scattering cross-section the position of the dip can be predicted.

\section{Elastic scattering in the generalized BB model}
\begin{figure}[H]
        \includegraphics[trim = 35mm 12mm 35mm 12mm, clip, width=0.8\linewidth]{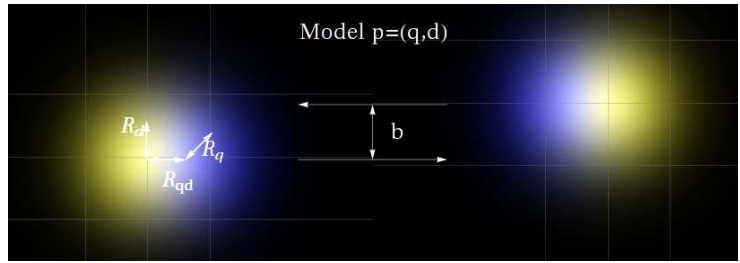}
        \centering
        \caption{(color online) Snapshot, illustration of the two scattering protons, when the proton is represented as a quark-diquark system,
        $p=(q,d)$ in the Bialas-Bzdak model. Both the quarks and the  diquarks are assumed to scatter as single entities, 
        corresponding to  Gaussians with radius $R_q$ and $R_{d}$, respectively. The quark is separated
	from the diquark by a distance of $R_{qd}$.
	All the model parameters follow a Gaussian distribution. 
	The impact parameter $b$ describes the 
	separation of the center of masses of the two colliding protons in the plane perpendicular to the direction of the beam.
	Based on Refs. \protect\cite{Bialas:2006qf,Bzdak:2007qq,Bialas:2006kw,Bialas:2007eg,Nemes:2012cp}.}
	\label{singlecoll}
\end{figure}

Proton-proton scattering is described in the BB model as a collision of two composite objects. Each of them is assumed to contain a constituent quark
and a diquark. The parametrization of the scattering situation is illustrated on Fig.~\ref{singlecoll}.
	In the original BB model the interaction between quarks and diquarks is assumed to be purely absorptive, 
        therefore the forward scattering amplitude of the BB model has no real part. In the present paper their formula is supplemented with 
	real part. This modification significantly improves the ability of the model to describe elastic $p+p$ scattering
	$d\sigma/dt$ in the dip region.

	The inelastic proton-proton cross-section in the impact parameter space for a fixed impact 
	parameter $\vec{b}$ is given by the following integral
    \begin{equation}
        \sigma_{\alpha}(b)=\int\limits^{+\infty}_{-\infty}...\int\limits^{+\infty}_{-\infty}{\text{d}^2s_q \text{d}^2s'_q \text{d}^2s_d \text{d}^2s'_d D(\vec{s_q},\vec{s_d})
        D(\vec{s_q}',\vec{s_d}')
        \sigma_{\alpha}(\vec{s_q},\vec{s_d};\vec{s_q}',\vec{s_d}';\vec{b})}\,, 
        \label{elsoegyenlet}
    \end{equation}
    	where $b=|\vec{b}|$ and the integral is taken over the two-dimensional transverse position vectors of the quarks $\vec{s_{q}}$, $\vec{s_{q}}'$ and  
	diquarks $\vec{s_{d}}$, $\vec{s_{d}}'$.

	The integral describes the convolution of the quark-diquark distributions of the
        two incoming protons with a $\sigma_{\alpha}$ function which in the original BB model provides the probability of inelastic interaction at
	given impact parameter vector $\vec{b}$ and at given quark, diquark transverse positions. In the present work this function is generalized
	to complex values, indicated by the subscript $\alpha$, with the more detailed motivation given after entailing the functional form of $\sigma_{\alpha}$.
        The complex valued $\sigma_{\alpha}$ function in the $\alpha=0$ case reduces to the real valued $\sigma(\vec{s_q},\vec{s_d};\vec{s_q}',\vec{s_d}';\vec{b})$ 
        function of the original BB model~\cite{Bialas:2006qf,Bzdak:2007qq,Bialas:2006kw,Bialas:2007eg,Nemes:2012cp}. The introduction of the
	$\alpha$ parameter is motivated by the Glauber-Velasco model \cite{Glauber:1984su,Glauber:1987sf,Glauber:2006gd}.

	Bialas and Bzdak in Ref.~\cite{Bialas:2006qf} supposed that the quark-diquark distribution of the proton follows a Gaussian shape
    \begin{equation}
        D\left(\vec{s_q},\vec{s_d}\right)=\frac{1+\lambda^2}{\pi R_{qd}^2}e^{-(s_q^2+s_d^2)/R_{qd}^2}\delta^2(\vec{s_d}+\lambda \vec{s_q}),\;\lambda=m_q / m_d\,,
    \end{equation}
	where $R_{qd}$ is the variance of the quark and diquark distance inside the proton and the $\lambda$ parameter
	is the mass ratio of the quark and diquark. The value of $\lambda=1/2$ would indicate a weakly bound diquark. The center
	of mass in the transverse plane is preserved with the help of a two-dimensional delta function.

	The original BB model supposes that protons are scattered elastically if and only if all of its constituents 
	are scattered elastically~\cite{Bialas:2006qf,Bzdak:2007qq,Glauber,Czyz:1969jg}
    \begin{align}
        \sigma(\vec{s_q},\vec{s_d};\vec{s_q}',\vec{s_d}';\vec{b})=1-\prod_{a,b \in \{q,d\}}\left[1-\sigma_{ab}(\vec{b} + \vec{s_a}' - \vec{s_b})\right]\,,
    \label{Glauber expansion}
    \end{align}
	where the inelastic differential cross-sections of the constituents are parametrized with Gaussian 
	distributions as well
    \begin{equation}
        \sigma_{ab}\left(\vec{s}\right) = A_{ab}e^{-s^2/R_{ab}^2},\;R_{ab}^2=R_a^2+R_b^2.
        \label{inelastic cross sections}
    \end{equation}

	A detailed study of the unmodified BB model, corresponding to the $\alpha=0$ case of our present work
	and also to a probabilistic interpretation
	of the forward scattering amplitude,
        has been completed  recently in Ref.~\cite{Nemes:2012cp}. It was shown that this model describes the main features of the $d\sigma/dt$
	data well in the ISR energy range of $\sqrt{s}=23-62$ GeV, except three data points around the dip of the $d\sigma/dt$ distribution,
        which were omitted from the optimalization~\cite{Nemes:2012cp}.

	Clearly, the original BB model neglects the real part of the forward scattering amplitude, 
        which results in $d\sigma/dt=0$ around the diffractive minimum, a feature that  does not correspond to the shape of 
	pp elastic scattering data.

 	From this point forward, a generalized version of the BB model is discussed
	which is able to describe the dip region of elastic scattering $d\sigma/dt$ at ISR energies, by modifying expression~(\ref{Glauber expansion}). 
	It takes into account, with a phenomenologically introduced parameter $\alpha$, that the proton is not always scattered elastically even if all of
	its constituents are scattered elastically. This generalized model is referred to as $\alpha$BB model. The $\alpha$BB model
  	is introduced with a purely imaginary factor in the formula
    \begin{align}
        \sigma_{\alpha}(\vec{s_q},\vec{s_d};\vec{s_q}',\vec{s_d}';\vec{b})=\left(1-i\alpha\right)\sigma(\vec{s_q},\vec{s_d};\vec{s_q}',\vec{s_d}';\vec{b}).
    \label{Glauber expansion with alpha}
    \end{align}

	The new parameter $\alpha$ is determined from the analysis of data. The $\alpha=0$ case corresponds to a situation, when 
	the proton always scatters elastically if its constituents scatter elastically. Parameter $\alpha$ is introduced in a way that is
	motivated by the $\alpha$ parameter of the Glauber-Velasco model of Refs. \cite{Glauber:1984su,Glauber:1987sf,Glauber:2006gd}. Parameter $\alpha$ can be considered
	as the $\rho$ parameter of parton-parton level scattering where $\rho$ is the ratio of the real to imaginary part of the forward scattering amplitude at
	zero momentum transfer. 

	From unitarity the elastic amplitude is the following
    \begin{equation}
        t_{el}(b)=i\left(1-\sqrt{1-\sigma_{\alpha}(b)}\right)\,,
	\label{telastic}
    \end{equation}
	where the elastic amplitude $t_{el}(b)$ is a complex valued function in our case; note that an imaginary unit is suppressed in the notation of the original BB model~\cite{Bialas:2006qf,Bzdak:2007qq,Bialas:2006kw,Bialas:2007eg}. Important
	to note also that the domain of the square root in Eq.~(\ref{telastic}) remains $\mathbb{C}$ even if $\alpha=0$.
	
    Recently, the important role of the real part of
    the elastic scattering amplitude in shaping $d\sigma/dt$ at the dip and in the Orear region was
    highlighted in \cite{Dremin:2012dm,Dremin:2012yh,Dremin:2012ke}.
    
In momentum transfer representation the amplitude of elastic scattering is given with the Fourier-transformation
        \begin{equation}
            T(\vec{\Delta})= \int\limits^{+\infty}_{-\infty}\int\limits^{+\infty}_{-\infty}{t_{el}(b)e^{i\vec{\Delta} \cdot \vec{b}}\text{d}^2b}=
            2\pi\int\limits_0^{+\infty}{t_{el}\left(b\right)J_0\left(\Delta b\right)b {\text d}b}\,,
        \end{equation}
    where $\Delta=|\vec{\Delta}|$ and $J_0$ is the zeroth Bessel-function of the first kind.

	Finally the differential cross-section is obtained as
    \begin{equation}
        \frac{d\sigma}{dt}=\frac{1}{4\pi}\left|T\left(\Delta\right)\right|^2\,.
    \end{equation}
	The real and
  	imaginary part combine together to describe the dip region of the differential cross section $d\sigma/dt$, even
	if $\alpha$ is only slightly different from 0. The results, as we shall also demonstrate below, will be qualitatively
	different from the BB model.
	
\subsection{Model $p=(q,d)$: The diquark is assumed to scatter as a single entity}

	This subsection is devoted to describe the remaining parts of the BB model when the diquark in 
	the proton is assumed to scatter as a single entity.
	In this case the two model parameters $A_{qd}$ and $A_{dd}$ can be expressed with the help of $A_{qq}$ if one
	supposes an idealized situation where the constituent diquark contains twice as many partons than the constituent quark without
	shadowing effects. This assumption may decrease the number of free parameters by two. The inelastic cross-sections
    \begin{equation}
    \label{totalinelastic}
        \sigma_{ab}=\int\limits^{+\infty}_{-\infty}\int\limits^{+\infty}_{-\infty}{\sigma_{ab}\left(\vec{s}\right)}\,\text{d}^2s= \pi A_{ab}R_{ab}^2,\;\; a,b \in \{q,d\}\,,
    \end{equation}
	are supposed to be proportional to the parton numbers. According to the assumption
    \begin{equation}
        \sigma_{qq}:\sigma_{qd}:\sigma_{dd}=1:2:4\,, 
        \label{ratiosforsigma}
    \end{equation}
	which corresponds to the case of no shadowing effect in the proton-proton scattering. Consequently
        \begin{equation}
            A_{qd}=A_{qq}\frac{4R_q^2}{R_q^2+R_d^2}\,,\;A_{dd}=A_{qq}\frac{4R_q^2}{R_d^2}\,.
        \end{equation}

	With these ingredients the computation of Eq.~(\ref{elsoegyenlet}) reduces to perform Gaussian integrals. Two
	of the Dirac delta functions simply express the transverse diquark coordinates with the corresponding quark coordinate
	using the mass ratio $\lambda$
    \begin{equation}
        \vec{s_d}=-\lambda \vec{s_q},\,\; \vec{s_d}'=-\lambda \vec{s_q}'.
    \end{equation}

	The remaining four Gaussian integrals lead to the following expression for each term in 
	the inelastic proton-proton cross-section formula (\ref{elsoegyenlet})
    \begin{equation}
        \frac{4v^2}{\pi^2}\int\limits^{+\infty}_{-\infty}\int\limits^{+\infty}_{-\infty}{\text{d}^2s_q \text{d}^2s_q' e^{-2v\left(s_q^2+s_q'^2\right)}
            \prod\limits^{k,l\,\in}_{\{q,d\}}e^{-c_{kl}\left(b-s_{k}+s_{l}'\right)^2}} =\frac{4v^2}{\Omega}e^{-b^2\frac{\Gamma}{\Omega}},
            \label{master_formula}
    \end{equation}

	where
    \begin{align}
        \Omega&=C_{qd,dq}\left[v+c_{qq} + \lambda^2 c_{dd}\right]+\left(1-\lambda\right)^2 D_{qd,dq}\,,\notag\\
        \Gamma&=C_{qd,dq}D_{qq,dd} + C_{qq,dd}D_{qd,dq}
            \label{master_formula1}
    \end{align}
	and
    \begin{equation}
	C_{kl,mn}=4v + \left(1+\lambda\right)^2\left(c_{kl}+c_{mn}\right)\,,\, D_{kl,mn}=v \left(c_{kl}+c_{mn}\right)+\left(1+\lambda\right)^2c_{kl}c_{mn}\,,
	            \label{master_formula2}
    \end{equation} 
     where the $c_{kl}$ parameters are abbreviations. 
	Formulas (\ref{master_formula}), (\ref{master_formula1}) and (\ref{master_formula2}) are equivalent with the corresponding formulas of
	the original BB model \cite{Bialas:2006qf}, although they are presented in a terser form.

\subsection{Model $p= (q,(q,q))$: The diquark is assumed to scatter as a composite object}
\begin{figure}[H]
        \includegraphics[trim = 35mm 12mm 35mm 12mm, clip, width=0.7\linewidth]{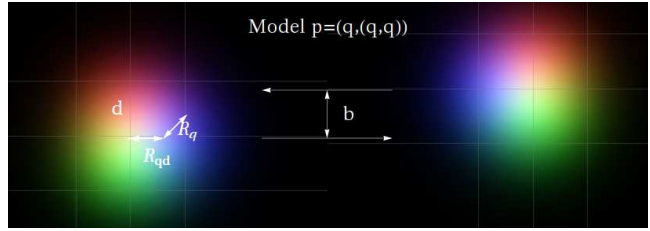}
        \centering
        \caption{Snapshot about the two scattering protons in the $p=(q,(q,q))$ model. Protons are described as a quark-diquark system, where the
        diquark is assumed to scatter as a quark-quark composite object. This is an illustration only, actually all the model parameters
	follow a Gaussian distribution, based on Refs. \protect\cite{Bialas:2006qf,Bzdak:2007qq,Bialas:2006kw,Bialas:2007eg,Nemes:2012cp}.} 
        \label{qqcoll}
\end{figure}

The collision of the protons in $p+p$ scattering is illustrated on Fig. \ref{qqcoll}, in case the diquark scatters as a composite object
including two constituent quarks.
 	Bialas and Bzdak \cite{Bialas:2006qf} supposed that the two quarks follow a Gaussian distribution inside the diquark
    \begin{equation}
            D\left(\vec{s_{q1}},\vec{s_{q2}}\right)=\frac{1}{\pi d^2}e^{-\left(s_{q1}^2+s_{q2}^2\right)/2d^2}
        \delta^2\left(\vec{s_{q1}}+\vec{s_{q2}}\right),
        \label{quarkdistribution}
    \end{equation}
 	where the transverse positions of the quarks are indicated with $\vec{s_{q1}}$, $\vec{s_{q2}}$. The distance $d$ is the RMS of the separation of the quarks
	inside the diquark. It is defined as
    \begin{equation}
        d^2=R_d^2-R_q^2,
    \end{equation}
	which expresses that the diquark is composed of two quarks which are separated by a distance $d$. In the present case the inelastic cross-sections $\sigma_{qd}$, $\sigma_{dq}$ and
	$\sigma_{dd}$ have to be factorized using the $\sigma_{qq}$ inelastic cross-section using expansion (\ref{Glauber expansion}). The
	formula for $\sigma_{qd}$ and $\sigma_{dq}$ is the following
    \begin{equation}
            \sigma_{qd}\left(\vec{s}\right)=\frac{4A_{qq}R^2_q}{R^2_d+R^2_q}e^{-s^2\frac{1}{R^2_d+R^2_q}}-\frac{A^2_{qq}R^2_{q}}{R^2_d}e^{-s^2/R^2_q},
    \end{equation}
	$\sigma_{dd}$ is more complicated, and is given in details in Refs. \cite{Bialas:2006qf,Bzdak:2007qq,Bialas:2006kw,Bialas:2007eg,Nemes:2012cp}. The inelastic cross-section (\ref{elsoegyenlet}) can be calculated using the master formula (\ref{master_formula})
	as before.

	Bialas and Bzdak originally used the total cross-section, the slope parameter $B$, the position of the dip and the position of the first diffractive maximum after
	the dip to determine the value of the fit parameters by solving 4 equations with 4 unknowns \cite{Bialas:2006qf}. They found, that the resulting parameters provide
        a good overall description of elastic scattering data at ISR energies. However, neither the errors of the parameters nor the fit quality description with a $\chi^2/NDF$ test were provided.  

	In Ref.~\cite{Nemes:2012cp} to determine the errors of the original BB model parameters a different method was applied, utilizing all the information in the data set by a
	multiparameter fit. In the present study we improved not only the fitting method, but also the BB model itself. We introduced the real part of the forward scattering amplitude with 
	parameter $\alpha$ motivated by the Glauber-Velasco model of \cite{Glauber:1984su,Glauber:1987sf}. This resulted in a new feature of the $\alpha$BB model, namely the ability to
	describe the data also around the diffractive minimum. Second, we have used the CERN MINUIT fitting package~\cite{James:1975dr} to determine the
	best values of the model parameters together with their errors. In the next section our results are presented, while we discuss the findings and compare them with earlier results in 
	the discussion part.

\section{Fit results in the 0.36 to 2.5 GeV$^2$ $\left|t\right|$ range}

	In this section the results of our MINUIT fits are presented for the ISR \cite{Nagy:1978iw,Amaldi:1979kd} and TOTEM \cite{Antchev:2011zz,Antchev:2011vs} proton-proton elastic
	scattering data. The scenario when the diquark is assumed to scatter as a single entity is considered first, which is followed by the fit results for the case
	when the diquark is considered as composite object. To provide a fair comparison among the model descriptions on the different dataset at different $\sqrt{s}$,
	the $\left|t\right|$ region of the fits is limited to the first TOTEM publication. In the discussion section, the fit quality is also studied in special fits to the TOTEM data in the
	low $\left|t\right|$ region. As we shall see, even the improved BB model cannot describe the TOTEM data in the whole $t$ region.

	The results show that thanks to the new parameter $\alpha$, which generates the real part of the forward scattering amplitude, the fits improve significantly and 
        describe the data also in the dip
	region, as compared to the $\alpha=0$ case presented in Refs \cite{Bialas:2006qf,Bzdak:2007qq,Bialas:2006kw,Bialas:2007eg,Nemes:2012cp}. This phenomenon can
	be interpreted such that the proton does not necessarily scatter elastically even if all its constituents are scattered elastically. This effect is small at lower
	collision energies, however it becomes, as we shall detail, more and more prominent with increasing energies.

\section{Model $p=(q,d)$: The diquark is assumed to scatter as a single entity}

\begin{figure}[H]
        \includegraphics[width=0.9\linewidth]{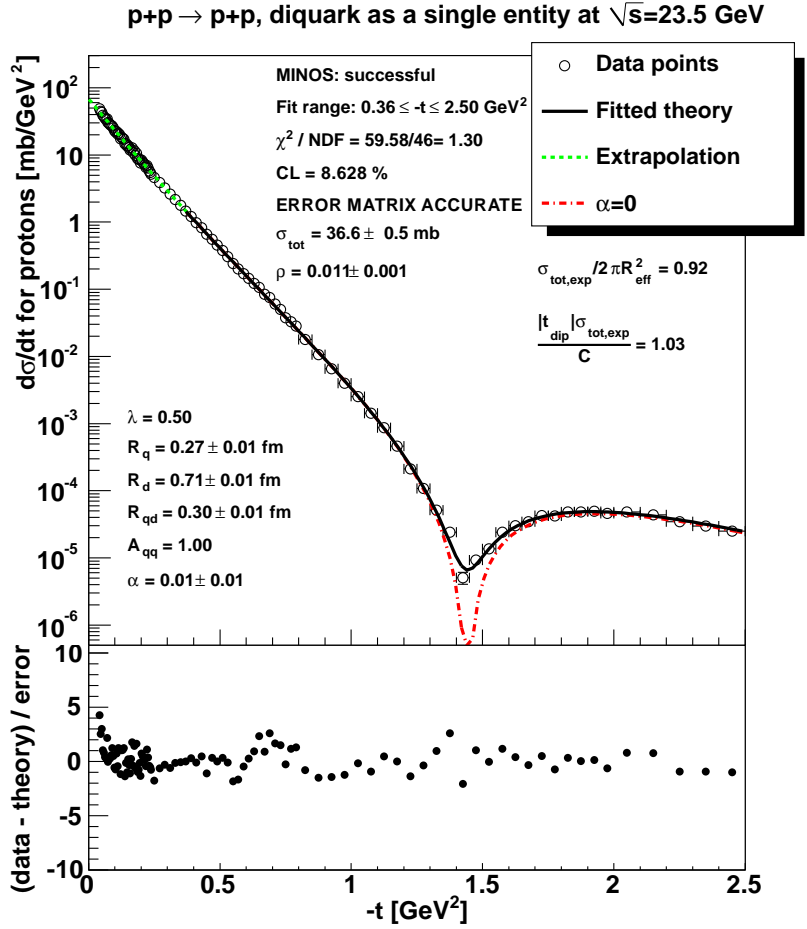}
        \centering
        \caption{Fit result at $\sqrt{s}=23.5$ GeV in case the diquark is assumed to be a single entity. Note that although $\alpha$ is not significantly different from
		$0$, a tiny value of $\alpha$ makes the fit behavior in the dip region significantly better, than the $\alpha=0$ case, indicated by a dashed line.
		The lower panel shows the deviation of the fitted theory (solid line) from the experimental data at that point. Note that below the legend we show also the
		test of two phenomenological relations which are discussed in the text. The total-cross section $\sigma_{tot,exp}$
		refers to the value of Ref.~\protect\cite{Antchev:2011vs}.
		}
	\label{single23}
\end{figure}

\begin{figure}[H]
        \includegraphics[width=0.9\linewidth]{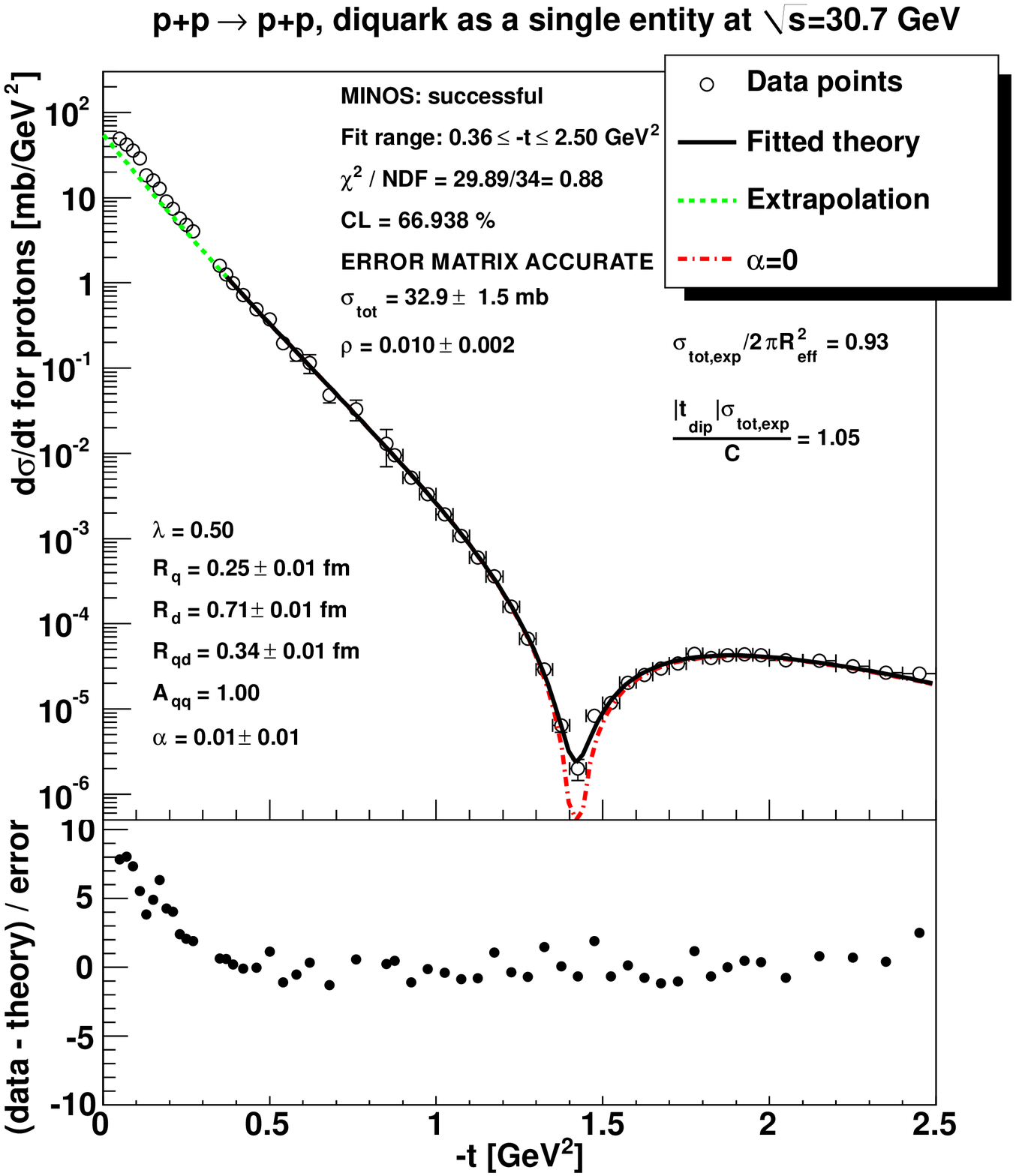}
        \centering
        \caption{Same as Fig. \ref{single23}, but at $\sqrt{s}=30.7$ GeV.}
	\label{single31}
\end{figure}

\begin{figure}[H]
        \includegraphics[width=0.9\linewidth]{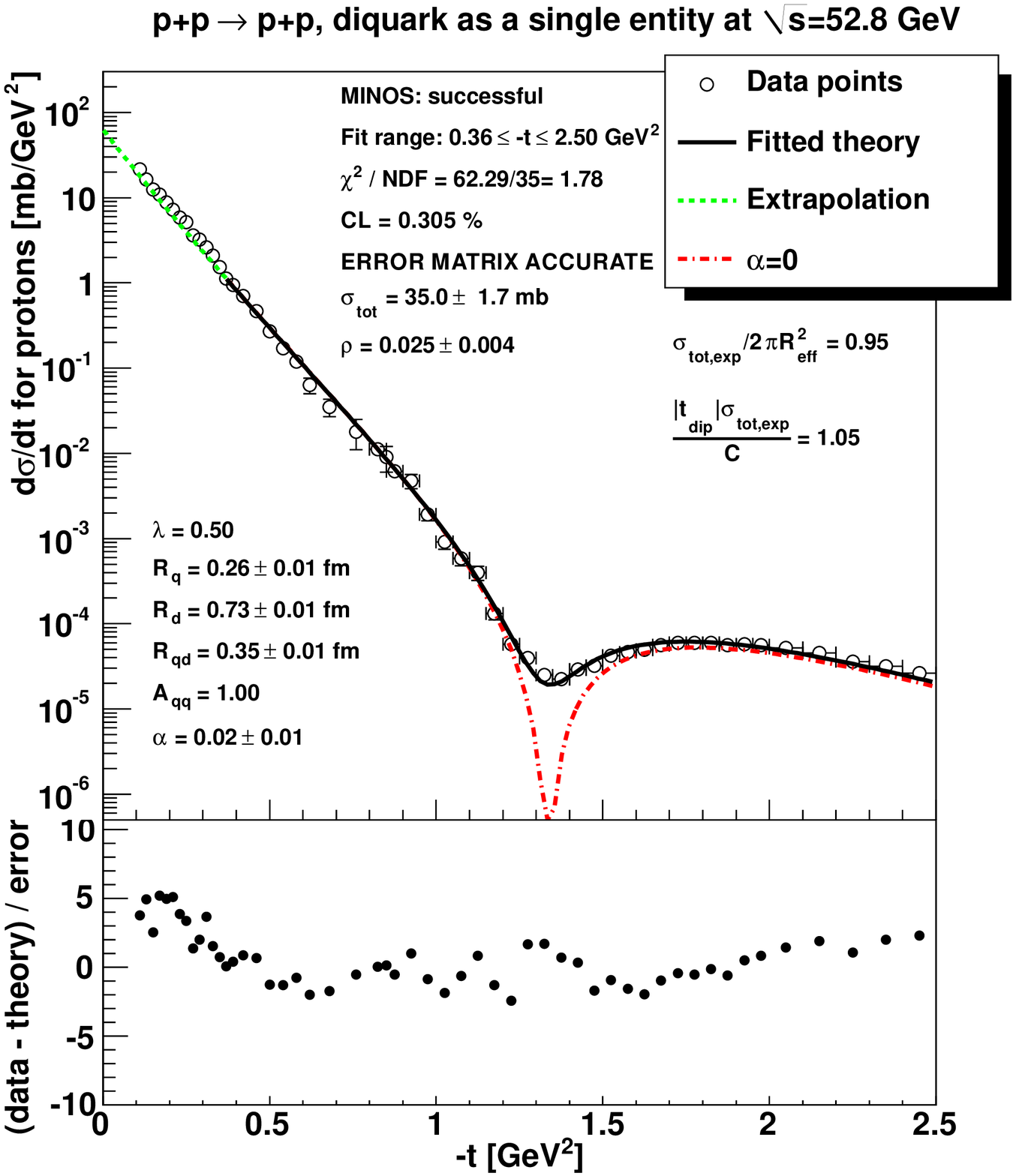}
        \centering
        \caption{Same as Fig. \ref{single23}, but at $\sqrt{s}=52.8$ GeV.}
	\label{single53}
\end{figure}

\begin{figure}[H]
        \includegraphics[width=0.9\linewidth]{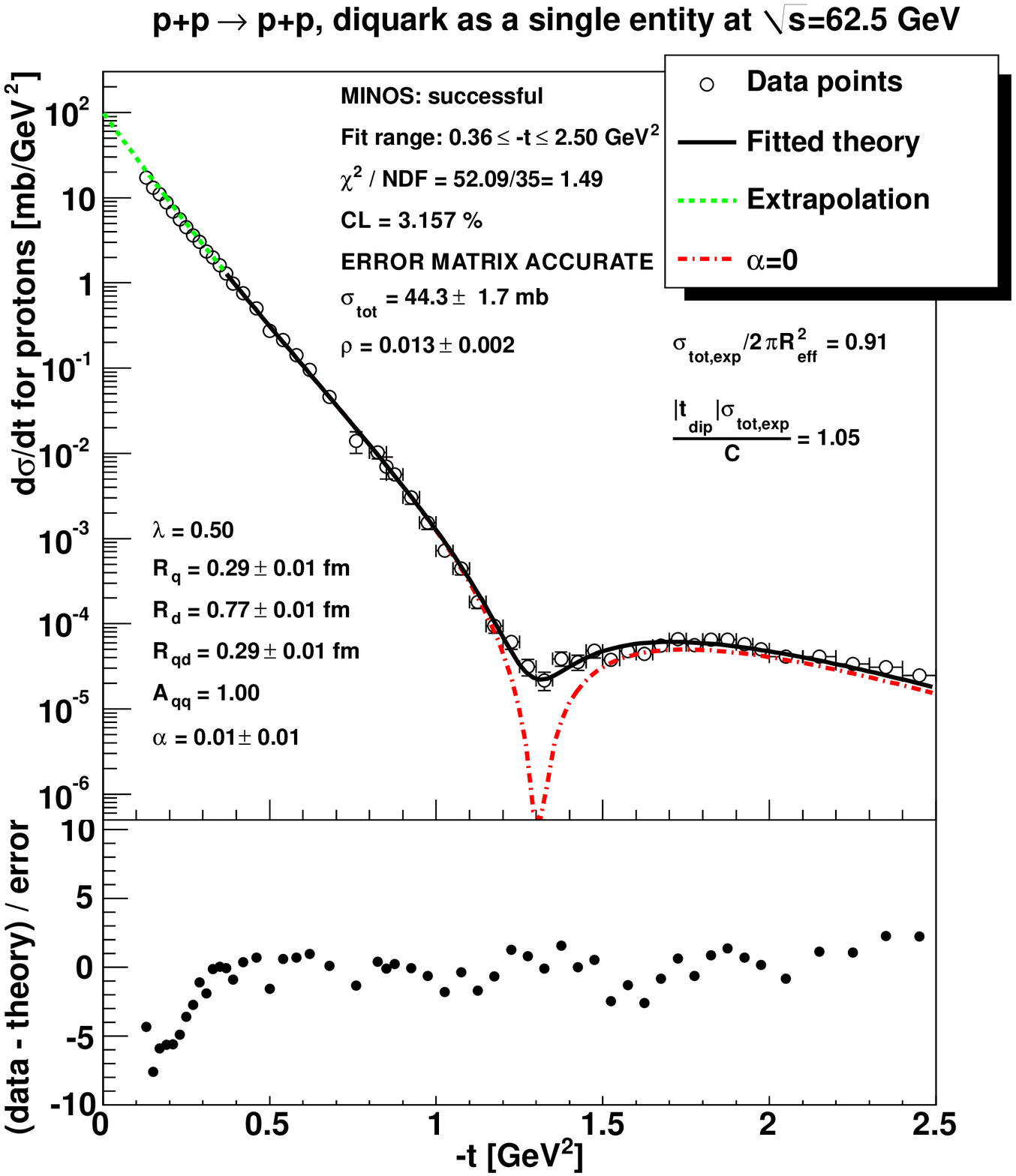}
	\centering
	\caption{Same as Fig. \ref{single23}, but at $\sqrt{s}=62.5$ GeV.}
	\label{single62}
\end{figure}

\begin{figure}[H]
        \includegraphics[width=0.9\linewidth]{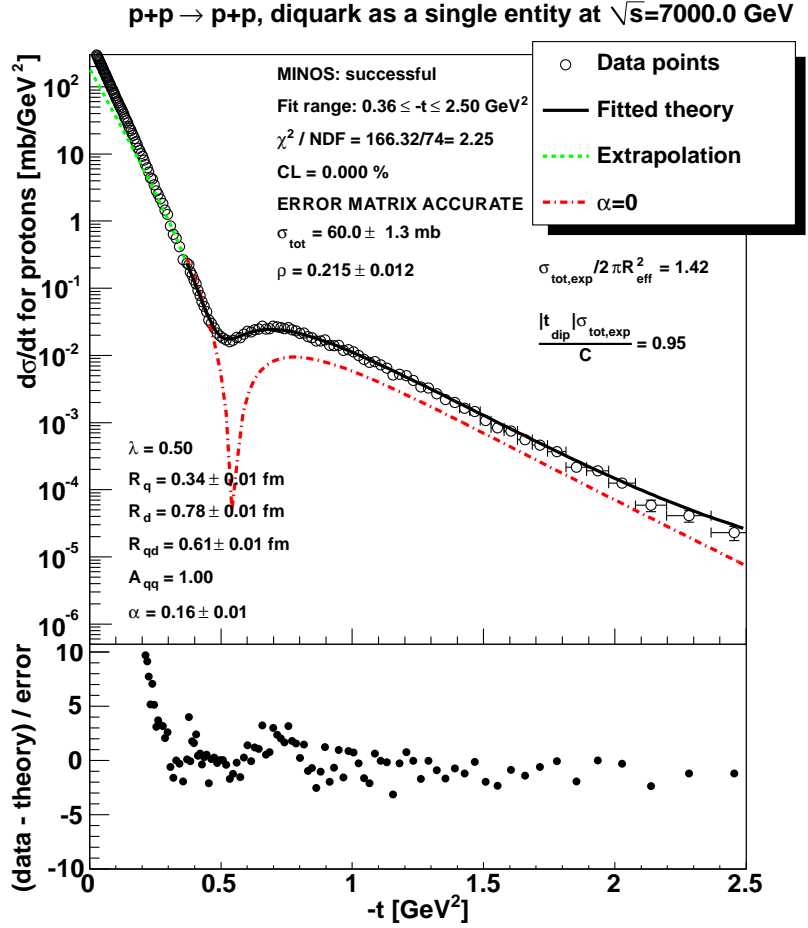}
	\centering
	\caption{Fit result at $\sqrt{s}=7$ TeV in case the diquark is assumed to interact as a single entity. Note
	that the $p=(q,d)$ version of the $\alpha BB$ model fails in the low-$\left|t\right|$ region, where $\sigma_{tot,exp}$
	of Ref.~\protect\cite{Antchev:2011vs} is underestimated. This may be the reason why
	$\frac{\sigma_{tot,exp}}{2 \pi R_{\mbox{\scriptsize \rm eff}}}$ is only 42\% close to 1.}
	\label{single7000}
\end{figure}

\begin{figure}[H]
	\centering
	\includegraphics[trim = 6mm 12mm 2mm 4mm, clip, width=0.4\linewidth]{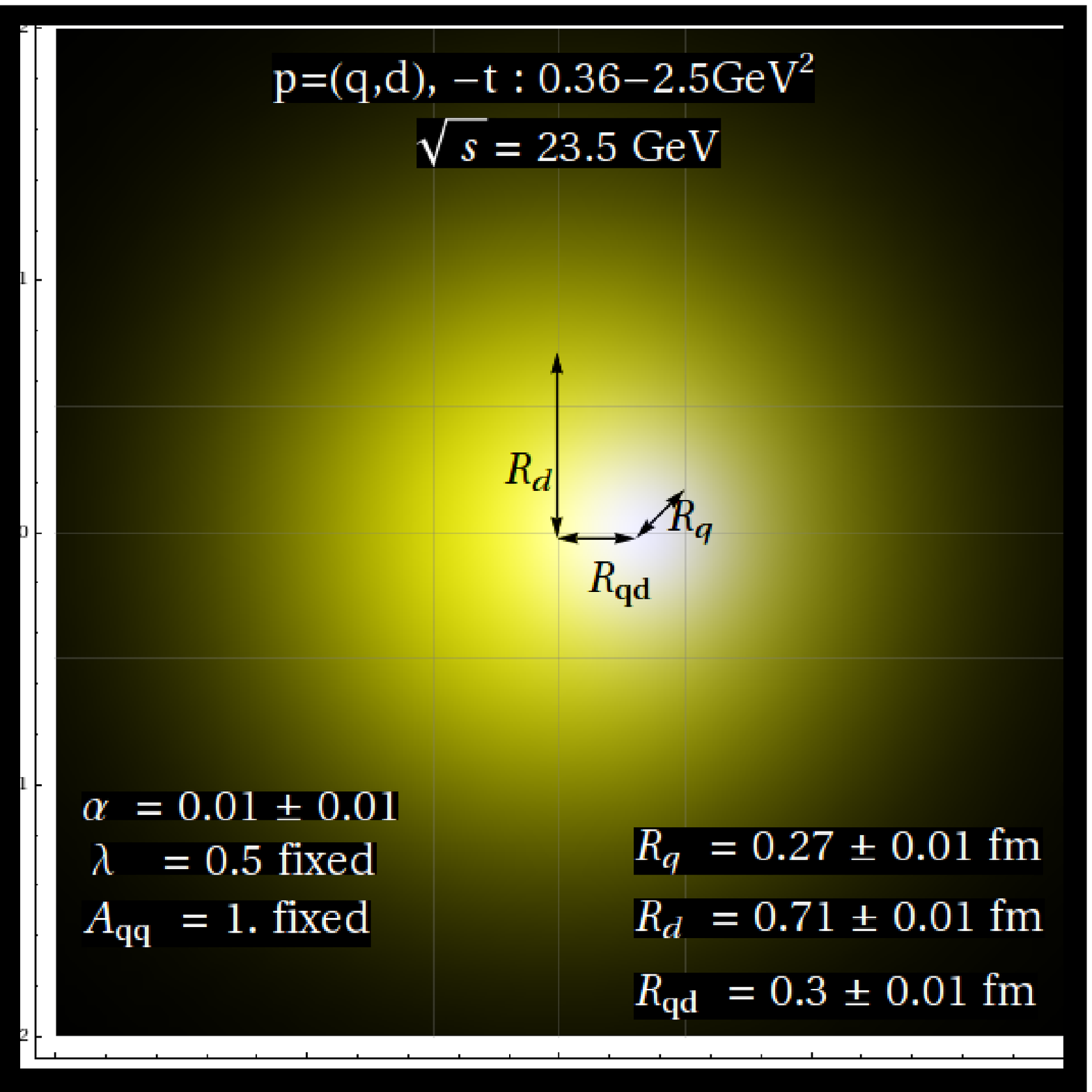}
	\includegraphics[trim = 6mm 12mm 2mm 4mm, clip, width=0.4\linewidth]{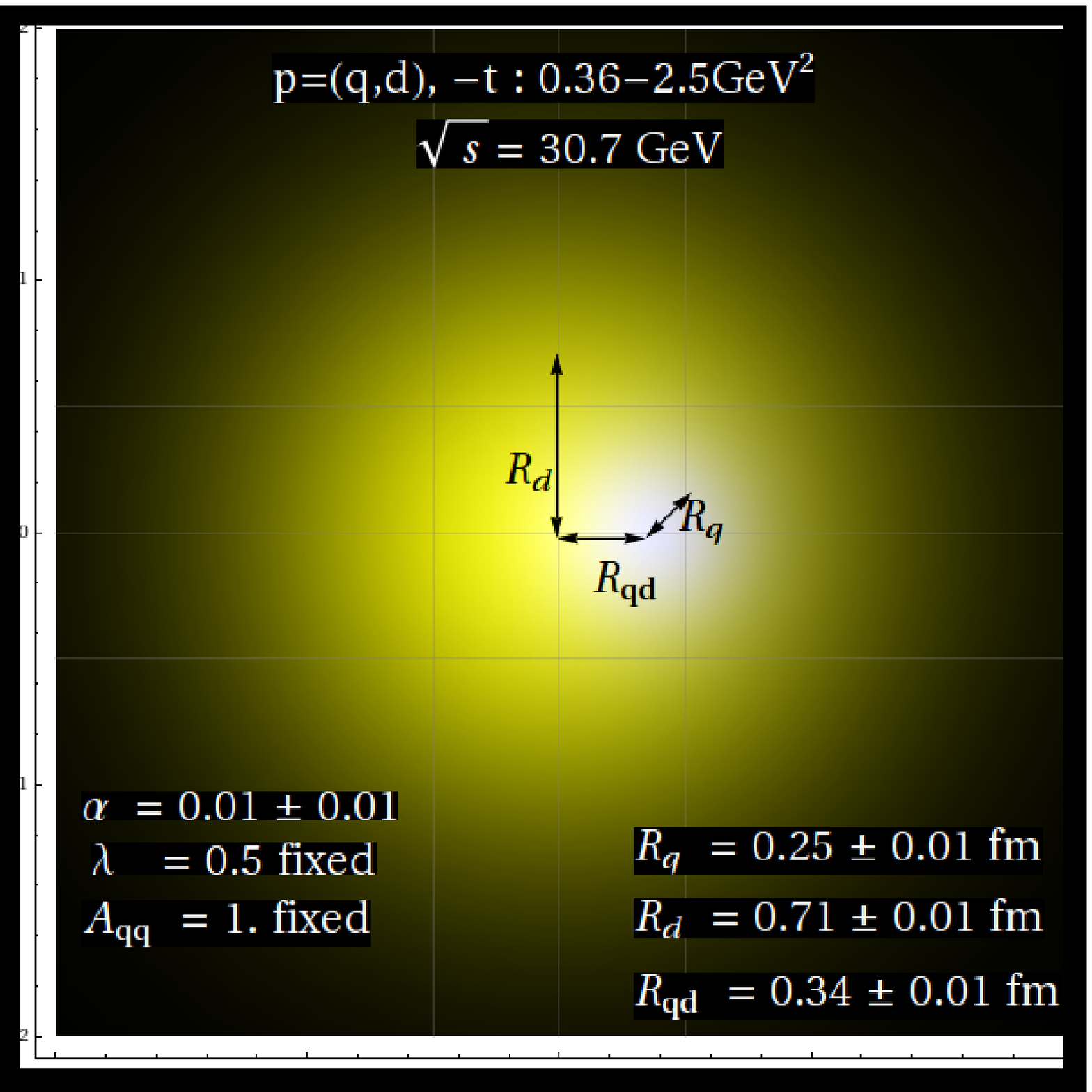}\\
	\includegraphics[trim = 6mm 12mm 2mm 4mm, clip, width=0.4\linewidth]{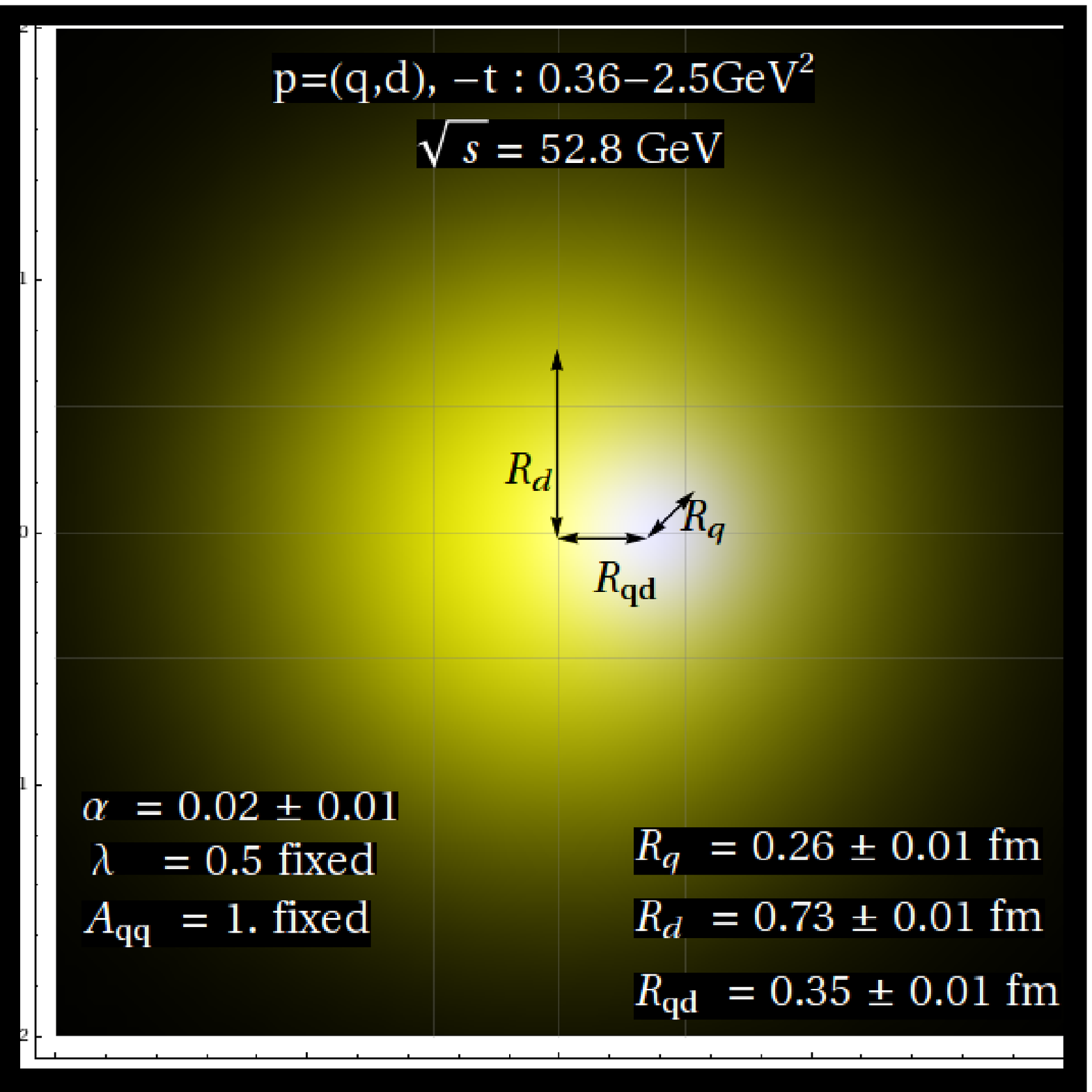}
	\includegraphics[trim = 6mm 12mm 2mm 4mm, clip, width=0.4\linewidth]{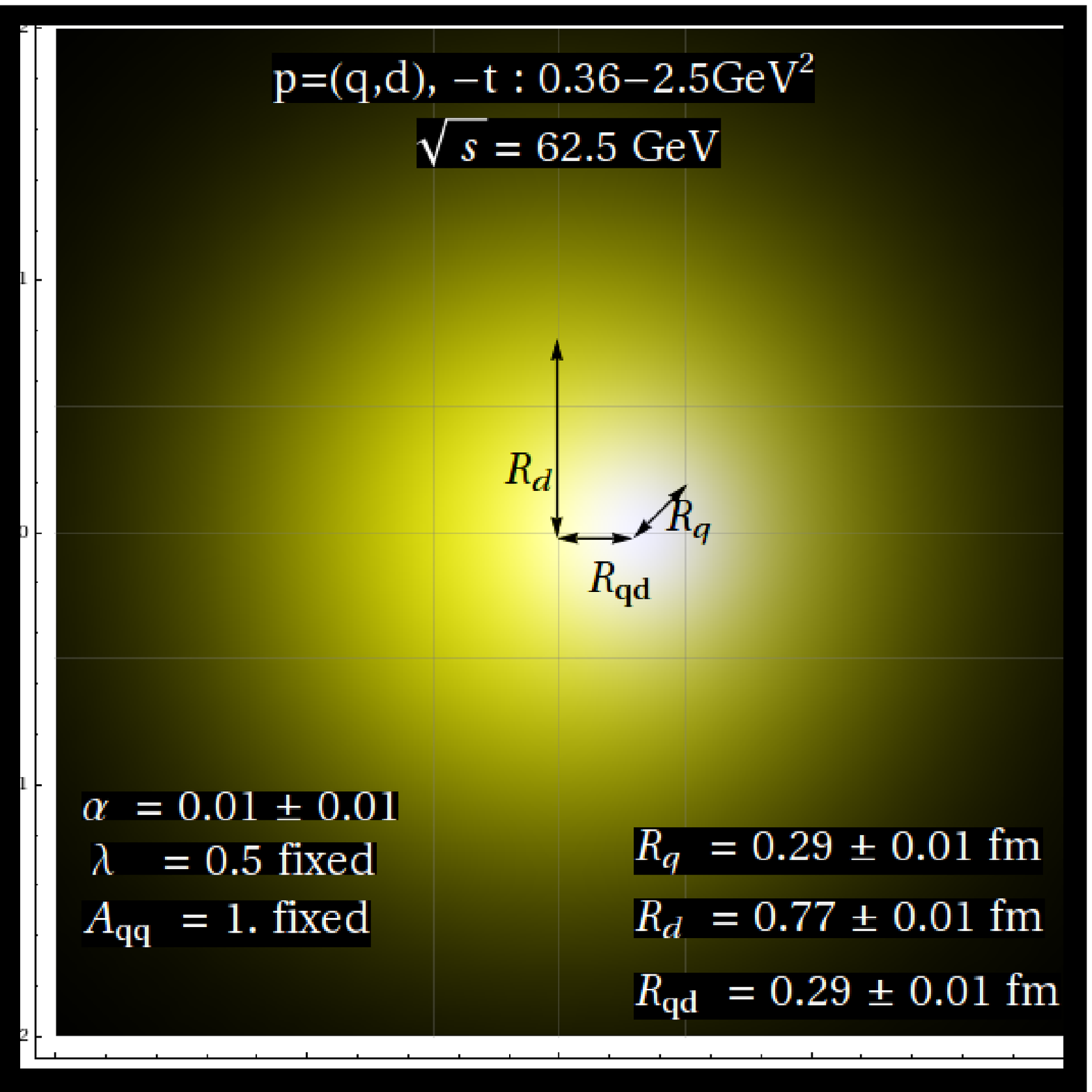}\\
      	\includegraphics[trim = 6mm 12mm 2mm 4mm, clip, width=0.4\linewidth]{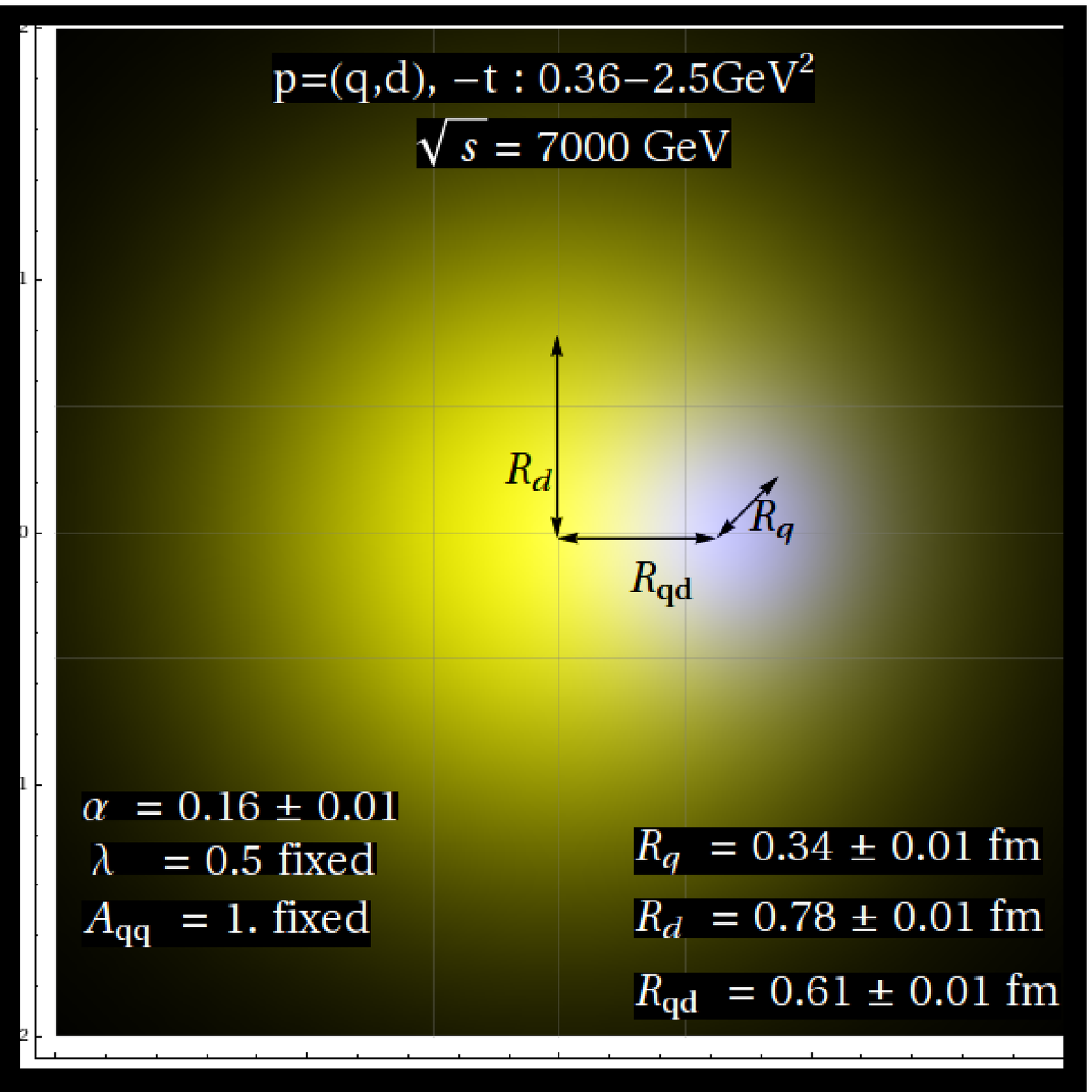}
	\caption{Visualization of the fit results in case the diquark is assumed to be a single entity.}
	\label{single_visualization}
\end{figure}

	In this subsection the fit results are collected in case the diquark is assumed to scatter as a single entity. This version of the model
	was fitted to the proton-proton elastic scattering data both at ISR \cite{Nagy:1978iw,Amaldi:1979kd} and at LHC	\cite{Antchev:2011zz} energy as well. The results are
	illustrated on Figs.~\ref{single23}-\ref{single7000} and a visualization of the model parameters is provided on Fig.~\ref{single_visualization}. The confidence levels, and model parameters with their errors are summarized in Table~\ref{tab:singleparameters}.

	The figures contain two phenomenological relations below the legend. The first relation suggests that the total-cross section is proportional to the area of a disk with
	an effective radius $R_{\mbox{\scriptsize \rm eff}}$, which is the square root of the quadratic sum of $R_q$, $R_d$ and $R_{qd}$
        \begin{align}
                R_{\mbox{\scriptsize \rm eff}} = \sqrt{R_{q}^2 + R_{d}^2 + R_{qd}^2}\;,
                \label{eqReff}
        \end{align}
        \begin{align}
                \sigma_{tot}=2 \pi R_{\mbox {\rm\scriptsize eff}}^2\;.
                \label{eqReffpersigmatot}
        \end{align}
	
	This formula was originally proposed for the BB model in \cite{Nemes:2012cp}, and was found to be model independently valid in both
	the $p=(q,d)$ and in the $p=(q,(q,q))$ models with a precision of $10-15\%$. In the current paper we validate this formula
	for the $\alpha\ne0$ case.
		
	Numerically, we have found another phenomenological formula, which indicates that the $|t|$ position of the first diffractive minimum $|t_{dip}|$ multiplied with the total-cross section
	$\sigma_{tot}$ is nearly constant
	\begin{equation} 
 		\frac{|t_{dip}|\,\sigma_{tot}}{C} \approx 1,
		\label{t_dip_sig_tot}
	\end{equation}
	where $C=54.8\pm0.7\,$mb GeV$^{2}$. The test of these relations for each energy with both the $p=(q,d)$ and $p=(q,(q,q))$ models is given on the figures and the results are collected and described in
	our Discussion part. 

\vfill
\eject

\begin{table}[!ht]
\centering
\begin{tabular}{|c|c|c|c|c|c|c|c|c|c|} \hline
$\sqrt{s}$ [GeV] & 23.5				& 30.7				& 52.8				& 62.5				& 7000					 \\ \hline\hline
$R_{q}$ [$fm$] &0.27 		 $\pm$ 0.01 	& 0.25 		 $\pm$ 0.01 	& 0.26 		 $\pm$ 0.01 	& 0.29 		 $\pm$ 0.01 	& 0.34 		 $\pm$ 0.01 	 	 \\ \hline
$R_{d}$ [$fm$] &0.71 		 $\pm$ 0.01 	& 0.71 		 $\pm$ 0.01 	& 0.73 		 $\pm$ 0.01 	& 0.77 		 $\pm$ 0.01 	& 0.78 		 $\pm$ 0.01 	 	 \\ \hline
$R_{qd}$ [$fm$] &0.30 		 $\pm$ 0.01 	& 0.34 		 $\pm$ 0.01 	& 0.35 		 $\pm$ 0.01 	& 0.29 		 $\pm$ 0.01	& 0.61 		 $\pm$ 0.01	 	 \\ \hline
$\alpha$ &0.01 		 $\pm$ 0.01 	& 0.01 		 $\pm$ 0.01 	& 0.02 		 $\pm$ 0.01 	& 0.01 		 $\pm$ 0.01 	& 0.16 		 $\pm$ 0.01 	 	 \\ \hline\hline
$\chi^2$/NDF & 59.58/46		& 	29.89/34	&  62.29/35	& 	52.09/35	& 	166.32/74		 \\   \hline
CL [\%] & 8.63				& 66.94				& 0.30				& 3.16				& 0.00					 \\   \hline
\end{tabular}
\caption{The $\sqrt{s}$ dependence of the confidence levels and the fit parameters in case the diquark is assumed to be a single entity, where the parameters $A_{qq}=1$ and $\lambda=0.5$ are fixed.
	The diquark size, as well as the inelasticity parameter $\alpha$ increase significantly with $\sqrt{s}$, indicating that the proton became larger and less elastic as $\sqrt{s}$ is increased
	to LHC energies.}
\label{tab:singleparameters}
\end{table}

\section{Model $p=(q,(q,q)$: diquark scatters as composite object}

\begin{figure}[H]
        \centering
        \includegraphics[width=0.9\linewidth]{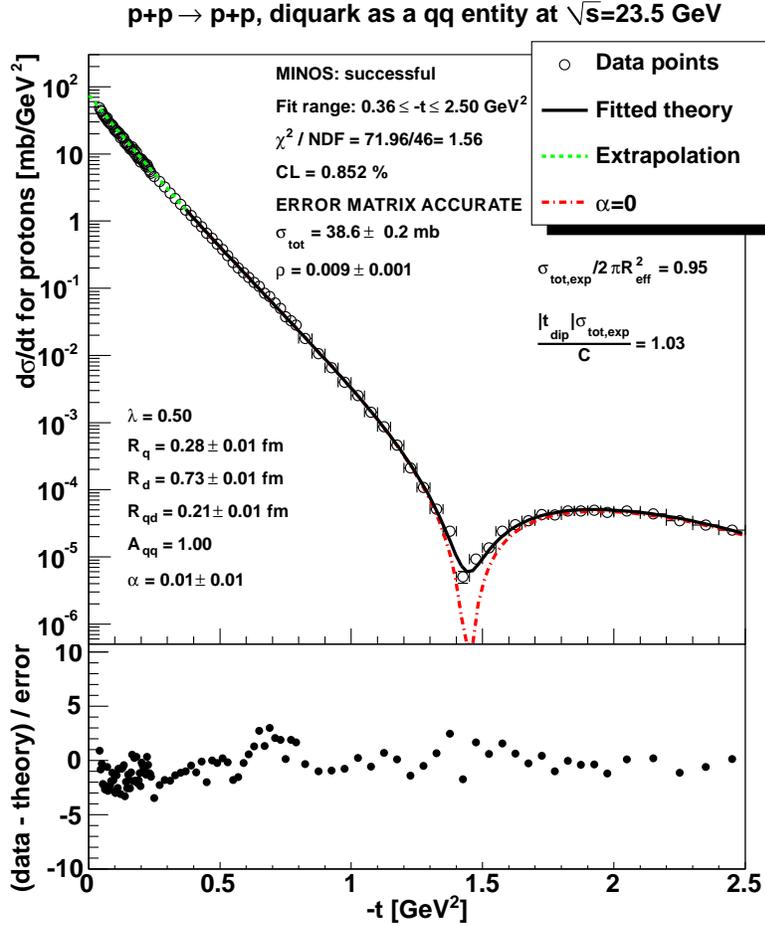}
        \caption{Fit result at $\sqrt{s}=23.5$ GeV in case the diquark is assumed to be a composite object. Note that even a small $\alpha\neq0$ value may make a big difference in
	the dip region. The fit becomes acceptable without a need to remove data points in this region. The fit quality parameters as well as two phenomenological relations are also
	indicated around the legend of the figure. Dashed line indicates the fit result when the parameter $\alpha$ is set to zero.}
	\label{qq23}
\end{figure}

\begin{figure}[H]
        \centering
        \includegraphics[width=0.9\linewidth]{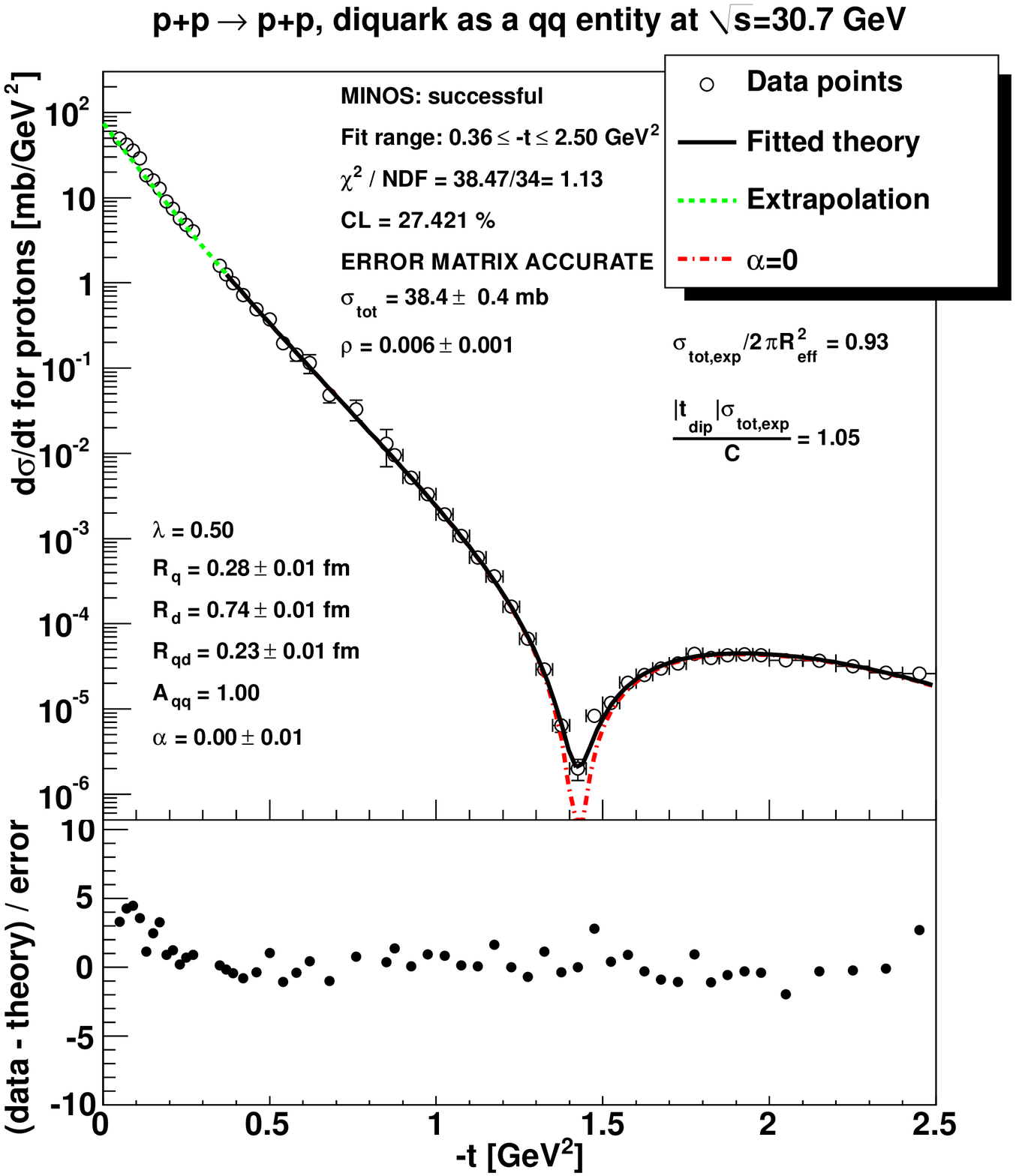}
        \caption{Same as Fig. \ref{qq23}, but at $\sqrt{s}=30.7$ GeV.}
	\label{qq31}
\end{figure}

\begin{figure}[H]
        \centering
        \includegraphics[width=0.9\linewidth]{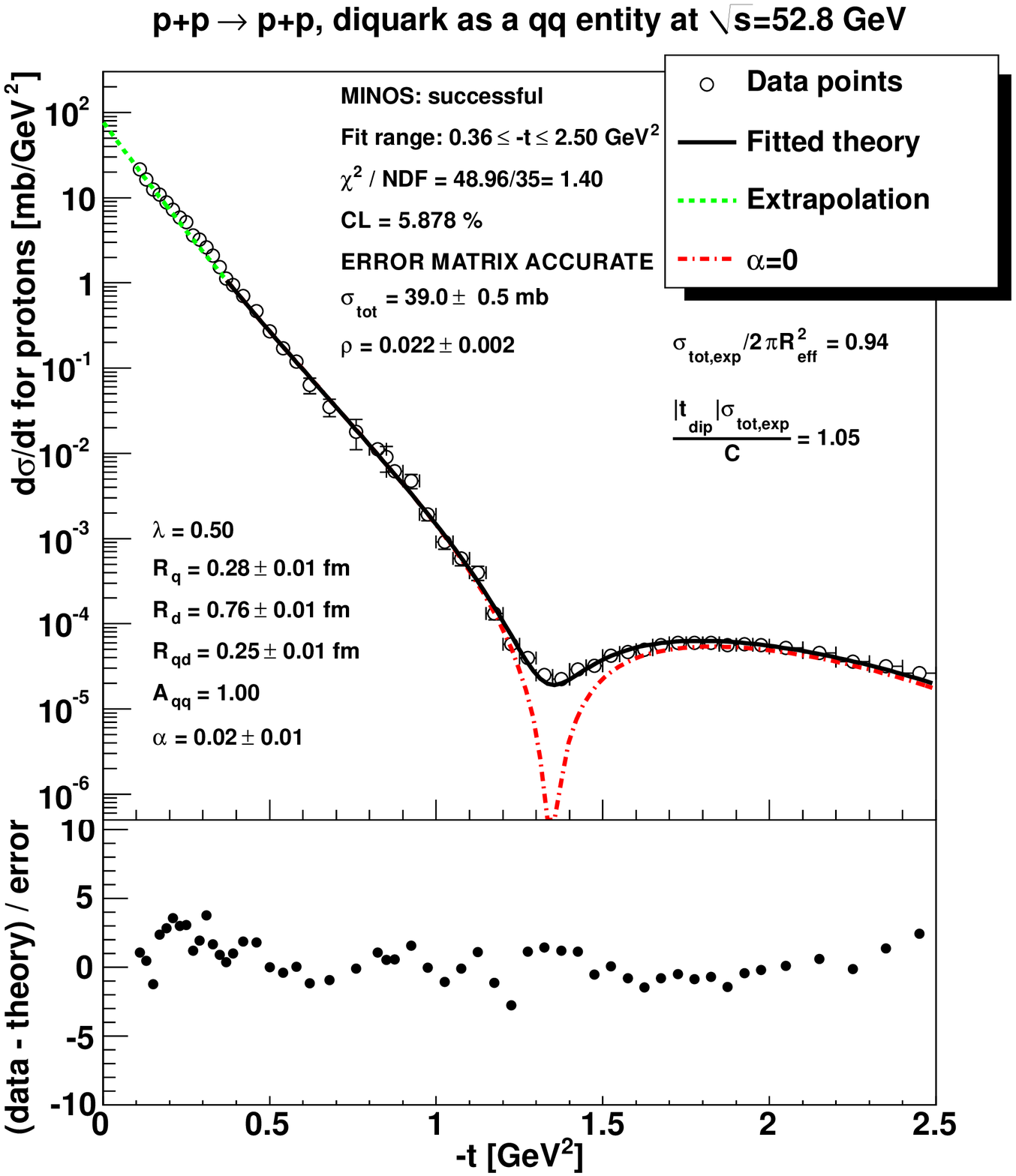}
        \caption{Same as Fig. \ref{qq23}, but at $\sqrt{s}=52.8$ GeV.}
	\label{qq53}
\end{figure}

\begin{figure}[H]
        \centering
        \includegraphics[width=0.9\linewidth]{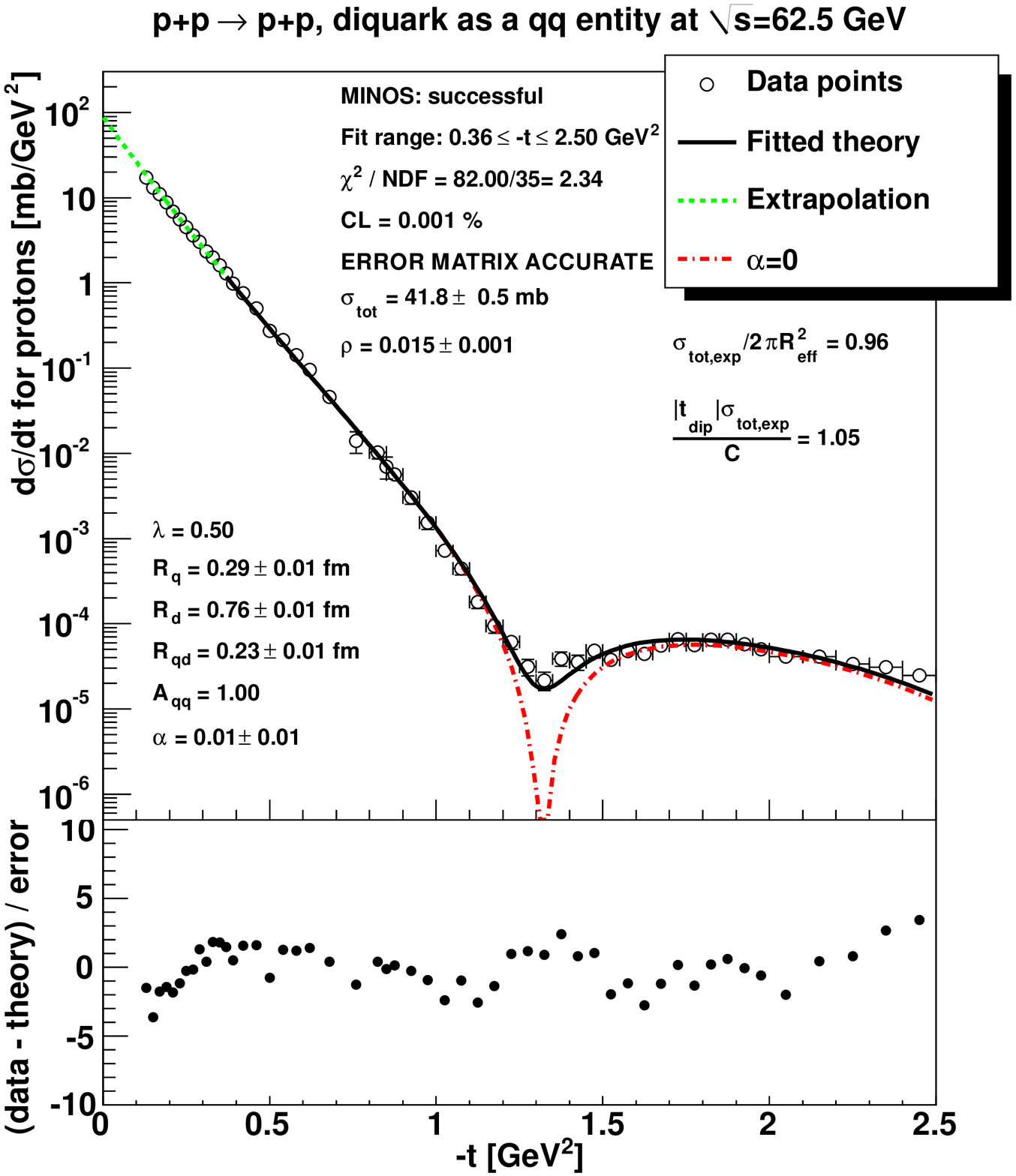}
	\caption{Same as Fig. \ref{qq23}, but at $\sqrt{s}=62.5$ GeV.}
	\label{qq62}
\end{figure}

\begin{figure}[H]
        \centering
        \includegraphics[width=0.9\linewidth]{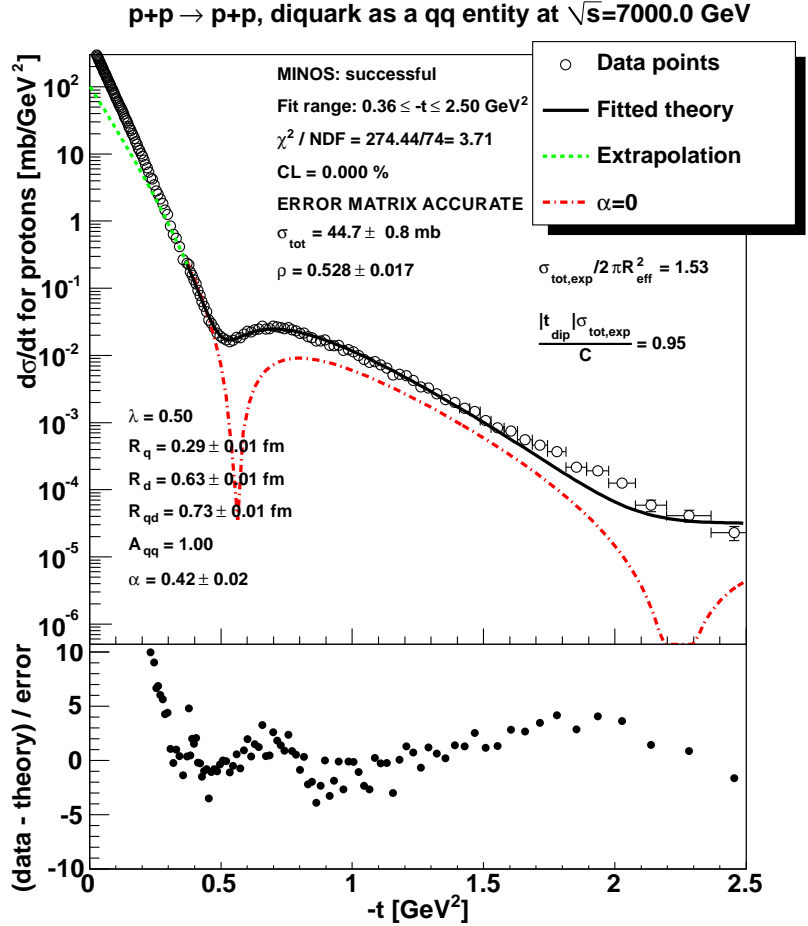}
	\caption{Same as Fig. \ref{qq23}, but at the LHC energy of $\sqrt{s}=7$ TeV. The extrapolation to low $\left|t\right|$ values indicates, that the 
	total cross-section is under estimated by the $\alpha$BB model, fitted in the intermediate $0.36\leq -t \leq 2.5$ GeV$^2$ region.
	Note
	that the $p=(q,(q,q))$ version of the $\alpha BB$ model fails in the low-$\left|t\right|$ region, where $\sigma_{tot,exp}$
	of Ref.~\protect\cite{Antchev:2011vs} is underestimated. This may be the reason why
	$\frac{\sigma_{tot,exp}}{2 \pi R_{\mbox{\scriptsize \rm eff}}}$ is only 53\% close to 1.}
	\label{qq7000}
\end{figure}

\begin{figure}[H]
        \centering
        \includegraphics[trim = 6mm 12mm 2mm 4mm, clip, width=0.4\linewidth]{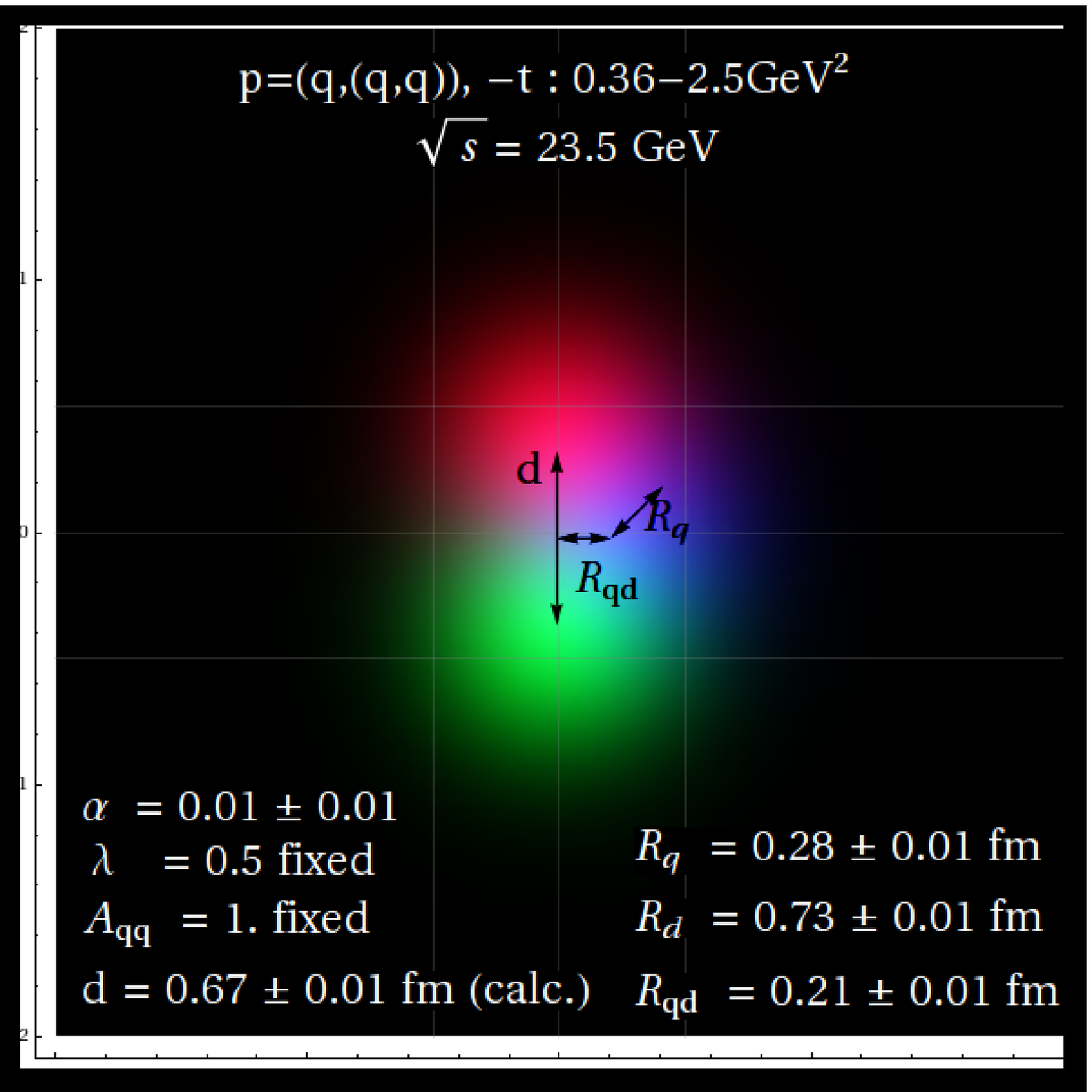}
        \includegraphics[trim = 6mm 12mm 2mm 4mm, clip, width=0.4\linewidth]{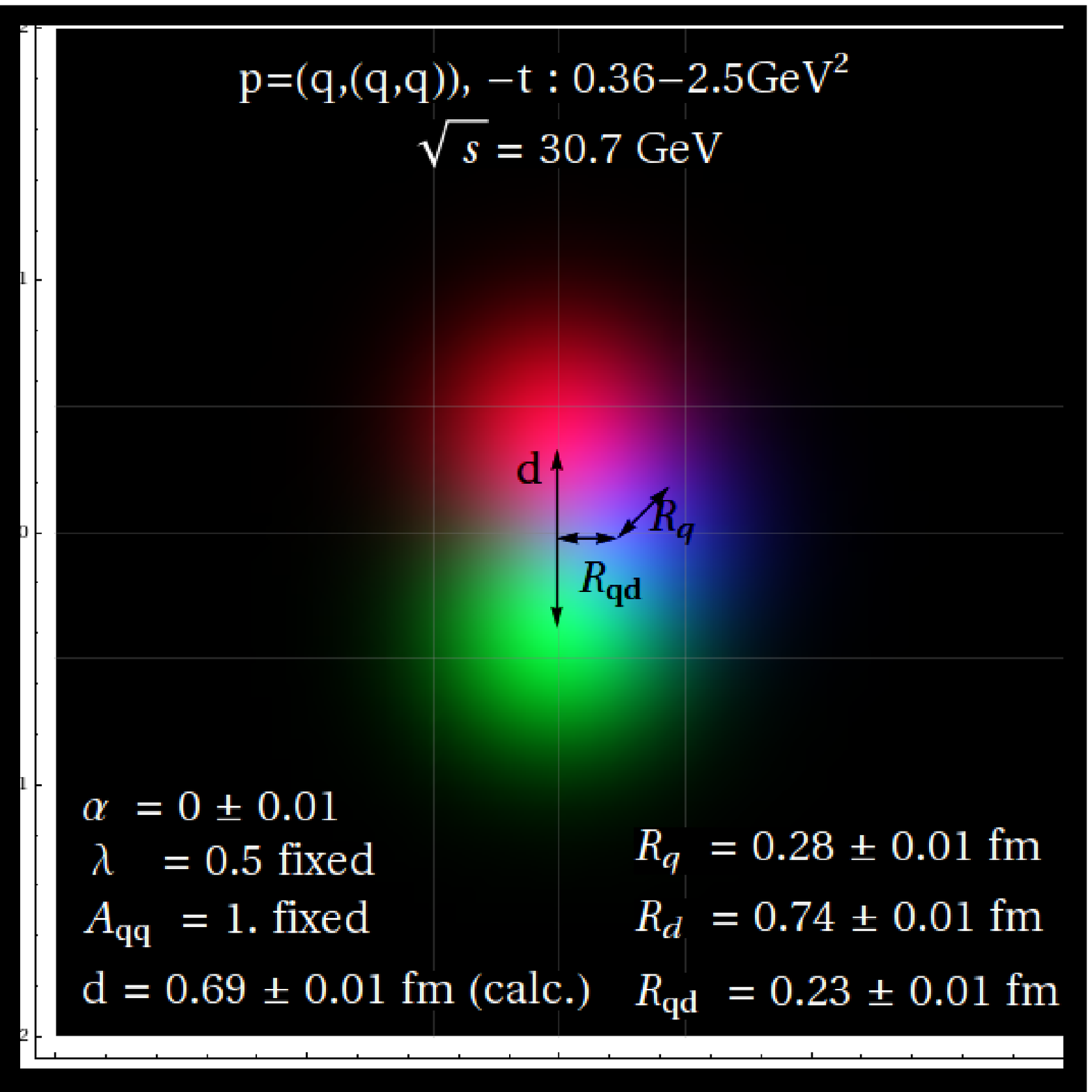}\\
        \includegraphics[trim = 6mm 12mm 2mm 4mm, clip, width=0.4\linewidth]{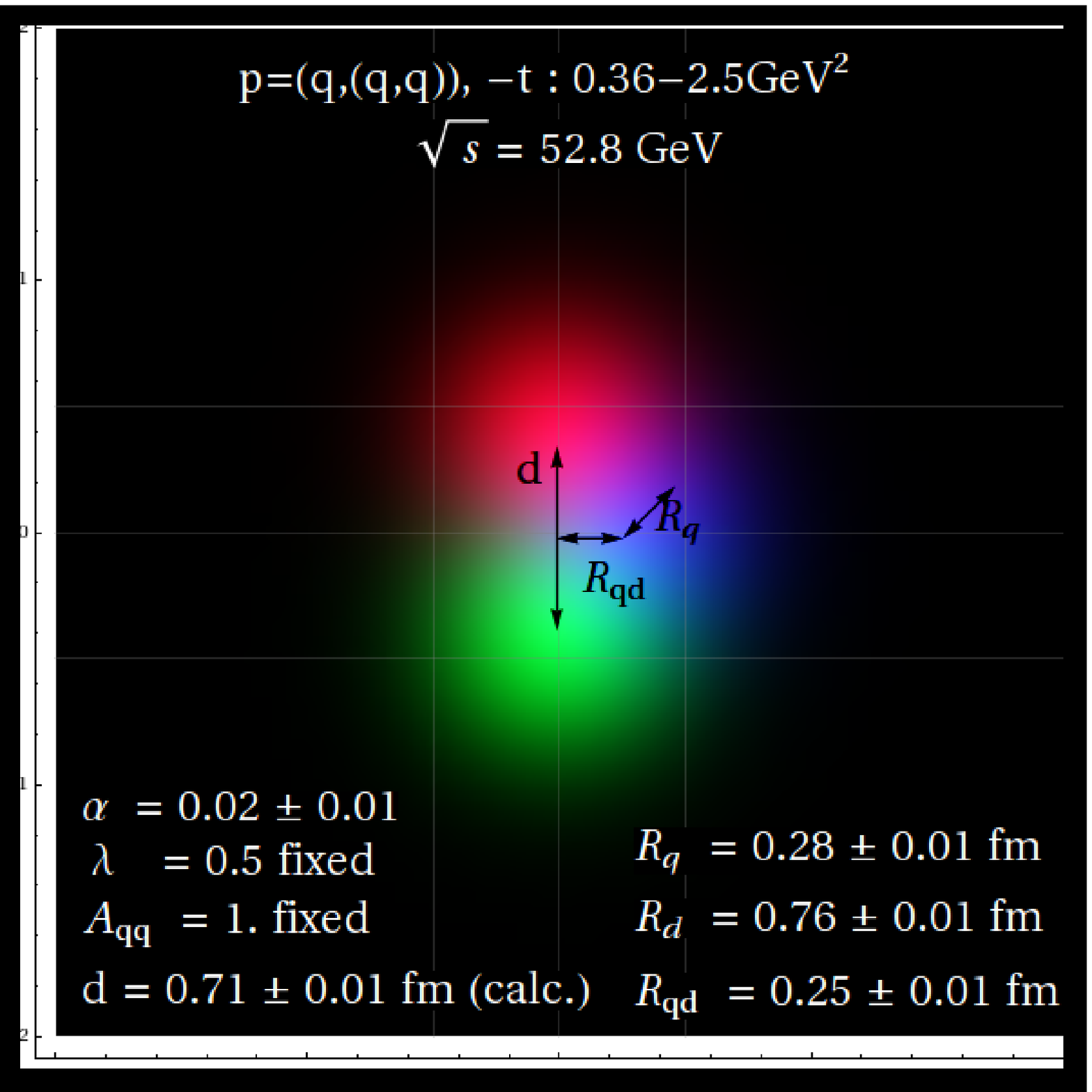}
        \includegraphics[trim = 6mm 12mm 2mm 4mm, clip, width=0.4\linewidth]{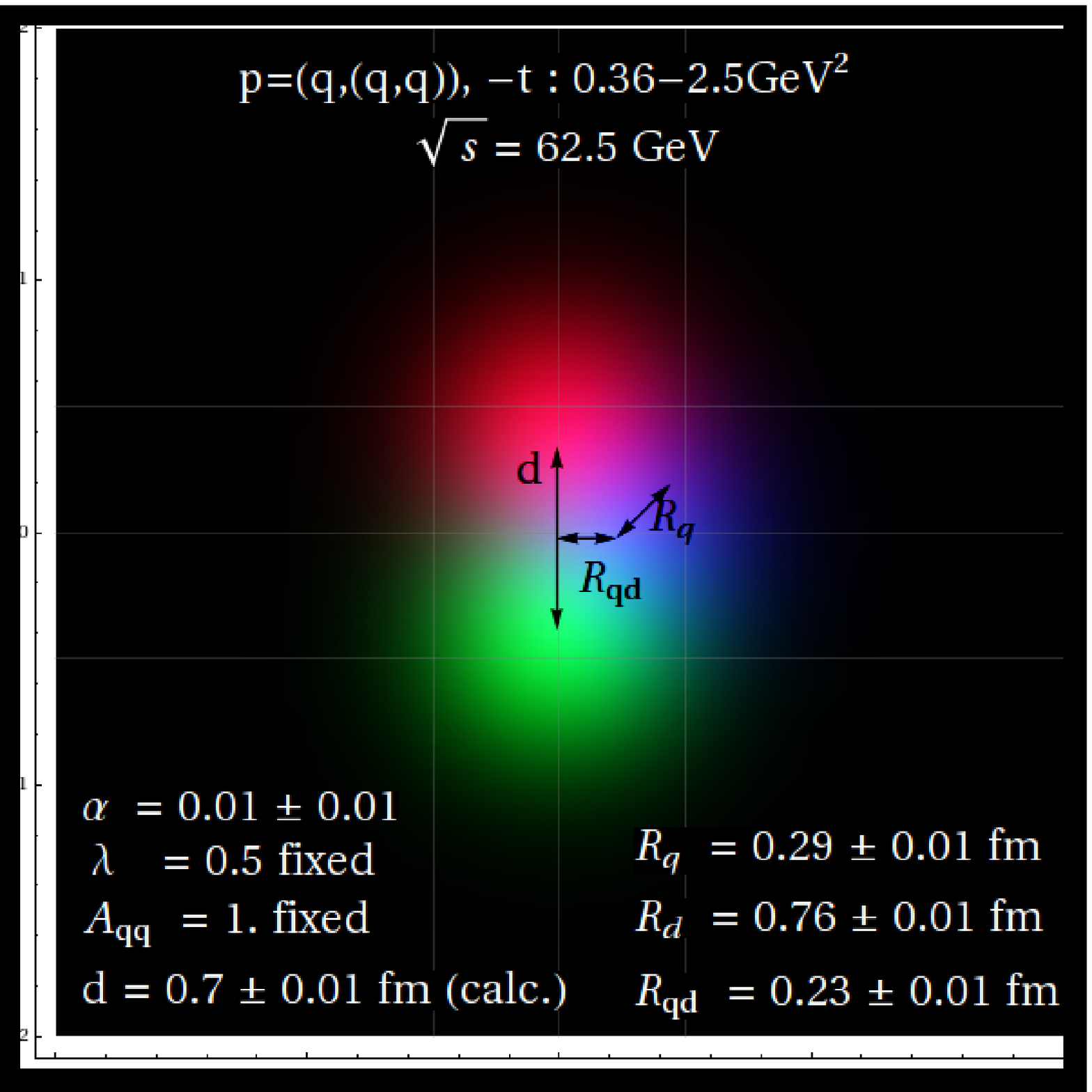}\\
        \includegraphics[trim = 6mm 12mm 2mm 4mm, clip, width=0.4\linewidth]{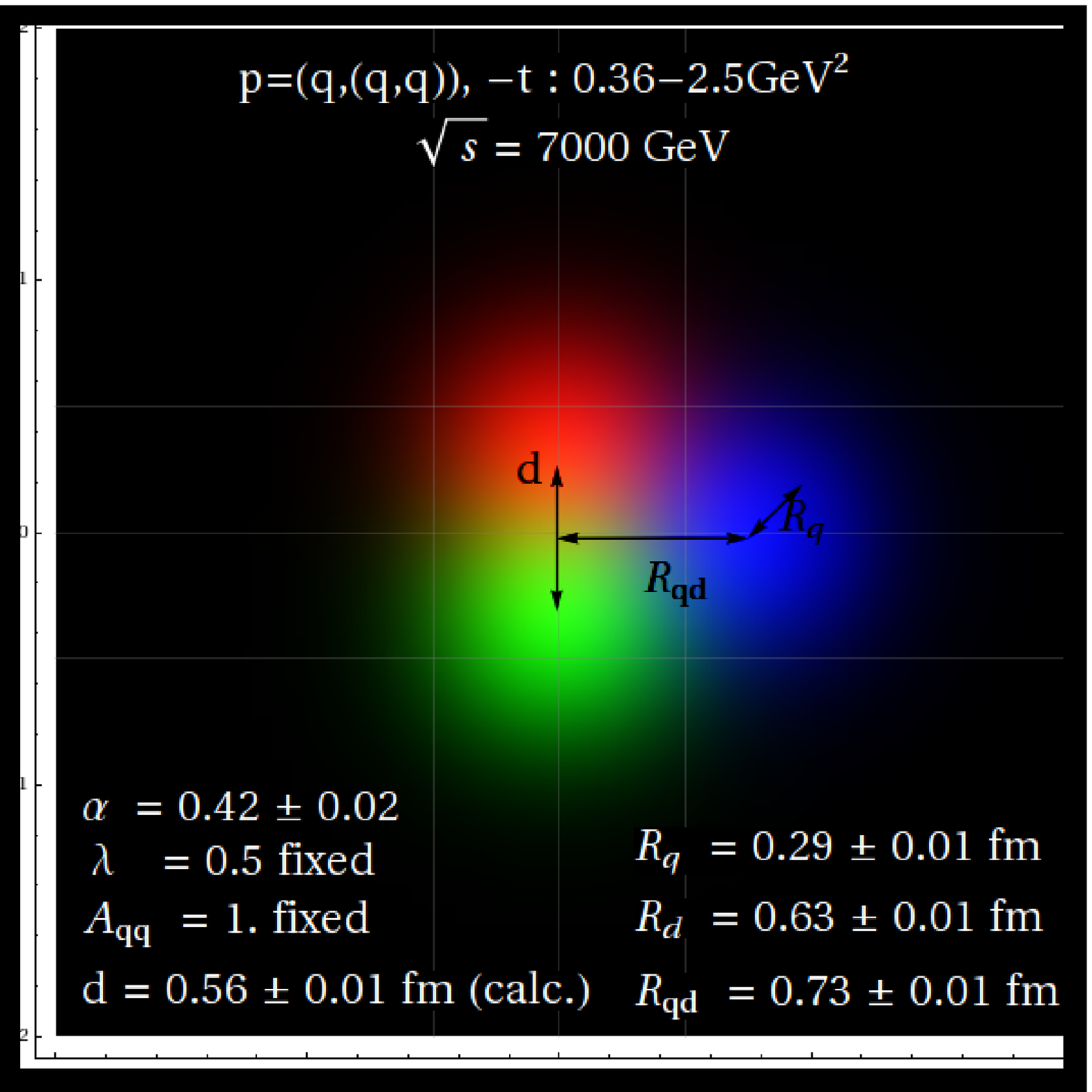}
	\caption{Visualization of the fit results in case the diquark is assumed to be a composite object.}
	\label{qq_visualization}
\end{figure}

The MINUIT fit results are presented as in the previous subsection, except that it is assumed that 
the diquark scatters as a composite object which contains two quarks. This version of the model
was fitted to the proton-proton elastic scattering data both at ISR \cite{Nagy:1978iw,Amaldi:1979kd} and at LHC
\cite{Antchev:2011zz} energy as well. The results are illustrated on Figs.~\ref{qq23}-\ref{qq7000}, while the visualization of the obtained parameters is given on Fig.~\ref{qq_visualization}. The confidence levels, and
model parameters with their errors are summarized in Table \ref{tab:qqparameters}.
\vfill
\eject

\begin{table}[!ht]
\centering
\begin{tabular}{|c|c|c|c|c|c|c|c|c|c|} \hline
$\sqrt{s}$ [GeV] & 23.5				& 30.7				& 52.8				& 62.5				& 7000					\\ \hline\hline
$R_{q}$ [$fm$] &0.28 		 $\pm$ 0.01 	& 0.28 		 $\pm$ 0.01 	& 0.28 		 $\pm$ 0.01 	& 0.29 		 $\pm$ 0.01 	& 0.29 		 $\pm$ 0.01 	 	\\ \hline
$R_{d}$ [$fm$] &0.73 		 $\pm$ 0.01 	& 0.74 		 $\pm$ 0.01 	& 0.76 		 $\pm$ 0.01 	& 0.76 		 $\pm$ 0.01 	& 0.63 		 $\pm$ 0.01 	 	\\ \hline
$R_{qd}$ [$fm$] &0.21 		 $\pm$ 0.01 	& 0.23 		 $\pm$ 0.01 	& 0.25 		 $\pm$ 0.01 	& 0.23 		 $\pm$ 0.01 	& 0.73 		 $\pm$ 0.01	 	\\ \hline
$\alpha$ &0.01 		 $\pm$ 0.01 	& 0.00 		 $\pm$ 0.01 	& 0.02 		 $\pm$ 0.01 	& 0.01 		 $\pm$ 0.01 	& 0.42 		 $\pm$ 0.02	 	 	\\ \hline\hline
$\chi^2$/NDF & 71.96/46		& 	38.47/34		& 	48.96/35			& 82.00/35	& 	274.44/74	 \\   \hline
CL [\%] &0.85				& 27.42				& 5.88				& 0.00				& 0.00					 \\   \hline
\end{tabular}
\caption{The $\sqrt{s}$ dependence of the confidence levels and the fit parameters in case the diquark is assumed to be a composite entity, where the parameters $A_{qq}=1$ and $\lambda=0.5$ are fixed. Note
	that the quark-diquark distance $R_{qd}$ increases significantly at the LHC, but note also that the fit quality
	is not statistically acceptable at 7 TeV.}
\label{tab:qqparameters}
\end{table}

\section{Total cross-section estimation based on TOTEM data in low $\left|t\right|$ range}

\begin{figure}[H]
        \centering
        \includegraphics[width=0.9\linewidth]{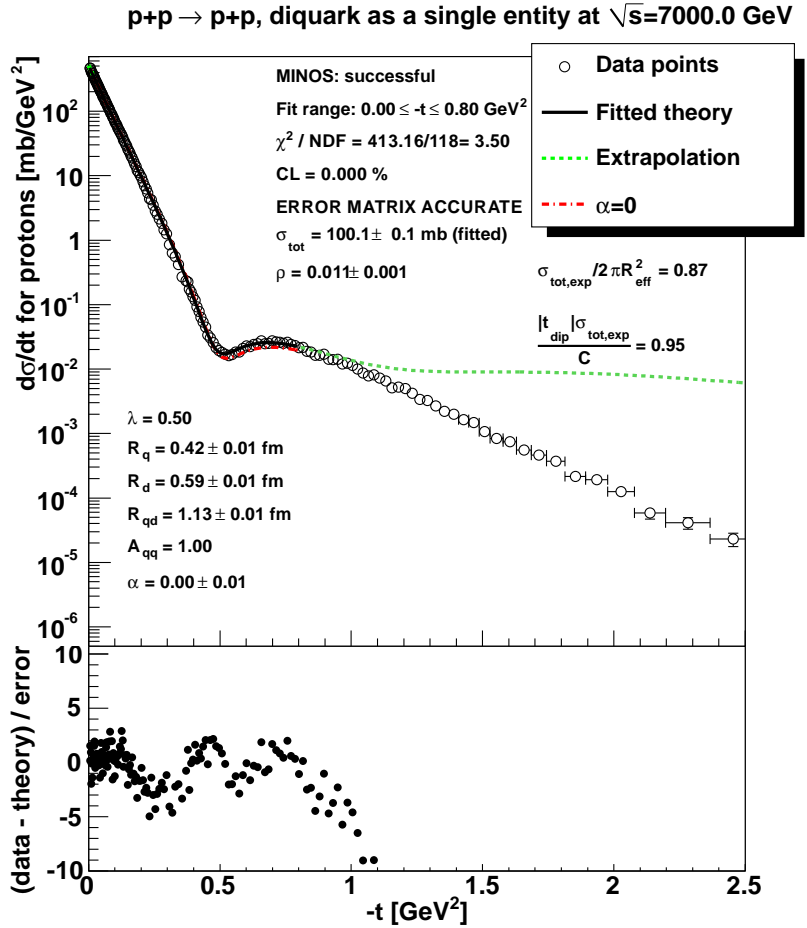}
	\caption{Fit result at $7$ TeV in the low $\left|t\right|$ range of $0.0 \leq -t \leq 0.8$ GeV$^2$, including the measured total cross-section value, as additional data point.
	The diquark is assumed to be a single entity.}
	\label{singlelow}
\end{figure}

\begin{figure}[H]
        \centering
        \includegraphics[width=0.9\linewidth]{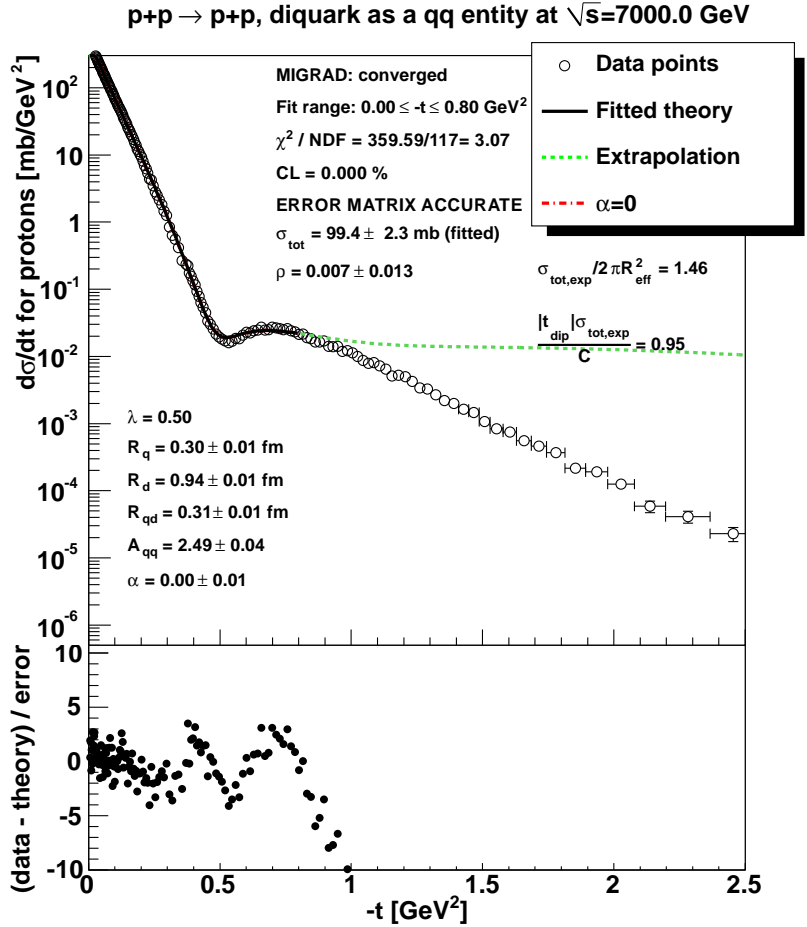}
	\caption{Same as Fig. \ref{singlelow}, but for the $p=(q,(q,q))$ model.}
	\label{qqlow}
\end{figure}

Note that even the generalized $\alpha$BB model failed on the TOTEM data when it was fitted in the $0.36\leq -t \leq 2.5$ GeV$^2$ region. In this section we investigate the fit range
dependence of this negative result. The total cross-section is estimated based on the small $\left|t\right|$ data which was
measured by the TOTEM experiment. The fit is repeated in a low $\left|t\right|$ range of $0.0 \leq -t \leq 0.8$ GeV$^2$ and the total cross-section value is \textit{included} into the fits as one
additional data point to show how well it can be described by the $\alpha$BB model. The results are illustrated in Figs.~\ref{singlelow} and~\ref{qqlow}.

In the low $\left|t\right|$-range a reasonable description can be achieved with the single entity version of the model, while surprisingly a less convincing result
is obtained when the diquark is assumed to be composite. However, when the fits are limited to the low $\left|t\right|$ region, the extrapolated fits deviate from the data
significantly in the large $\left|t\right|$ region for both the $p=(q,d)$ and the $p=(q,(q,q))$ model.

Note also that we could not find a fit with statistically acceptable quality, when the fit region was increased to $0.0 \leq -t \leq a$ GeV$^2$, where $a>0.8$ GeV$^2$. This
indicates that the $\alpha$BB model does not describe TOTEM elastic scattering data in the whole measured $t$ range.
\vfill
\eject

\section{Discussion}

\begin{figure}[H]
	\subfigure[]{\label{without_sigtot},\includegraphics[width=0.48\linewidth]{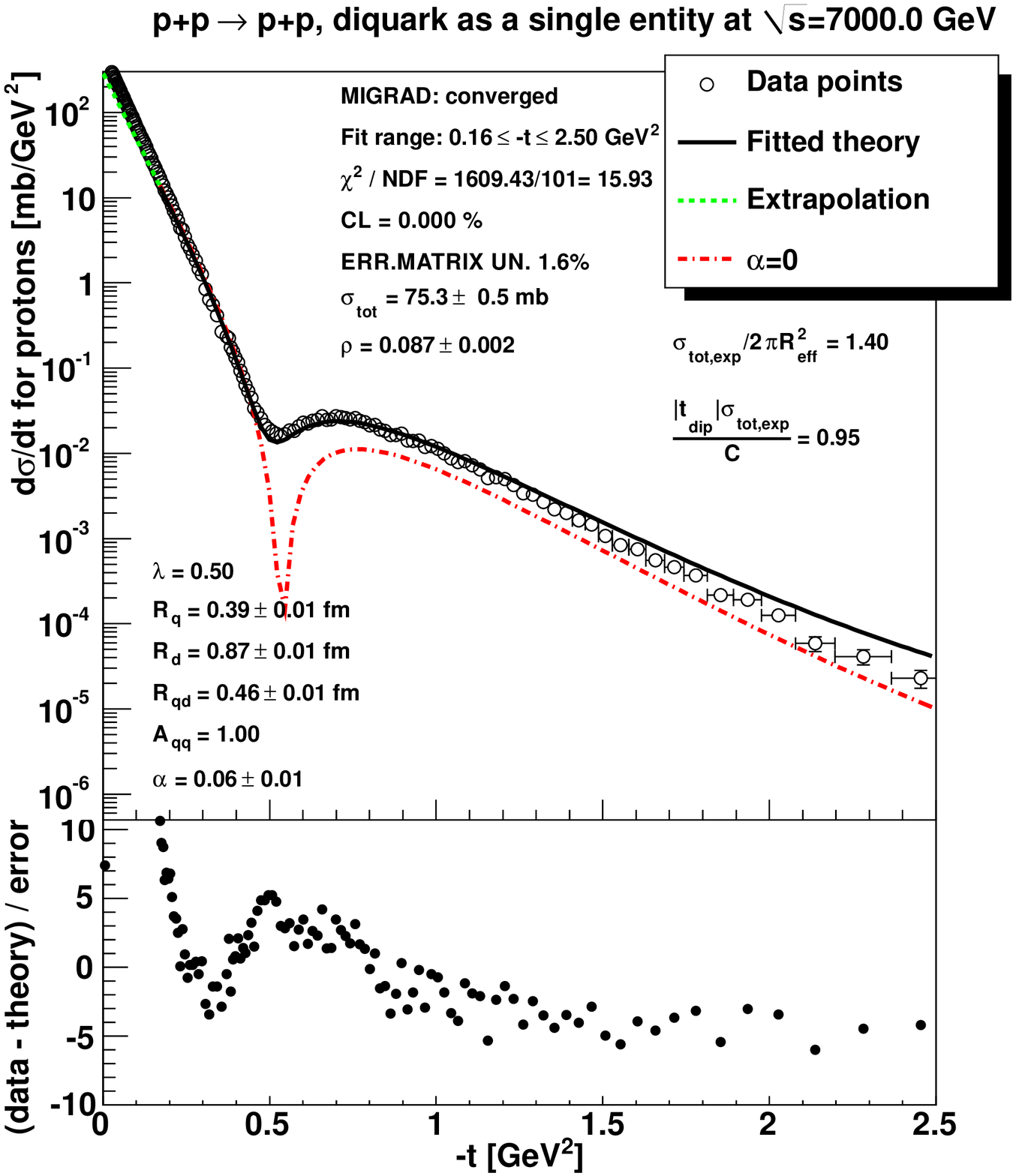}}
        \subfigure[]{\label{with_sigtot},\includegraphics[width=0.48\linewidth]{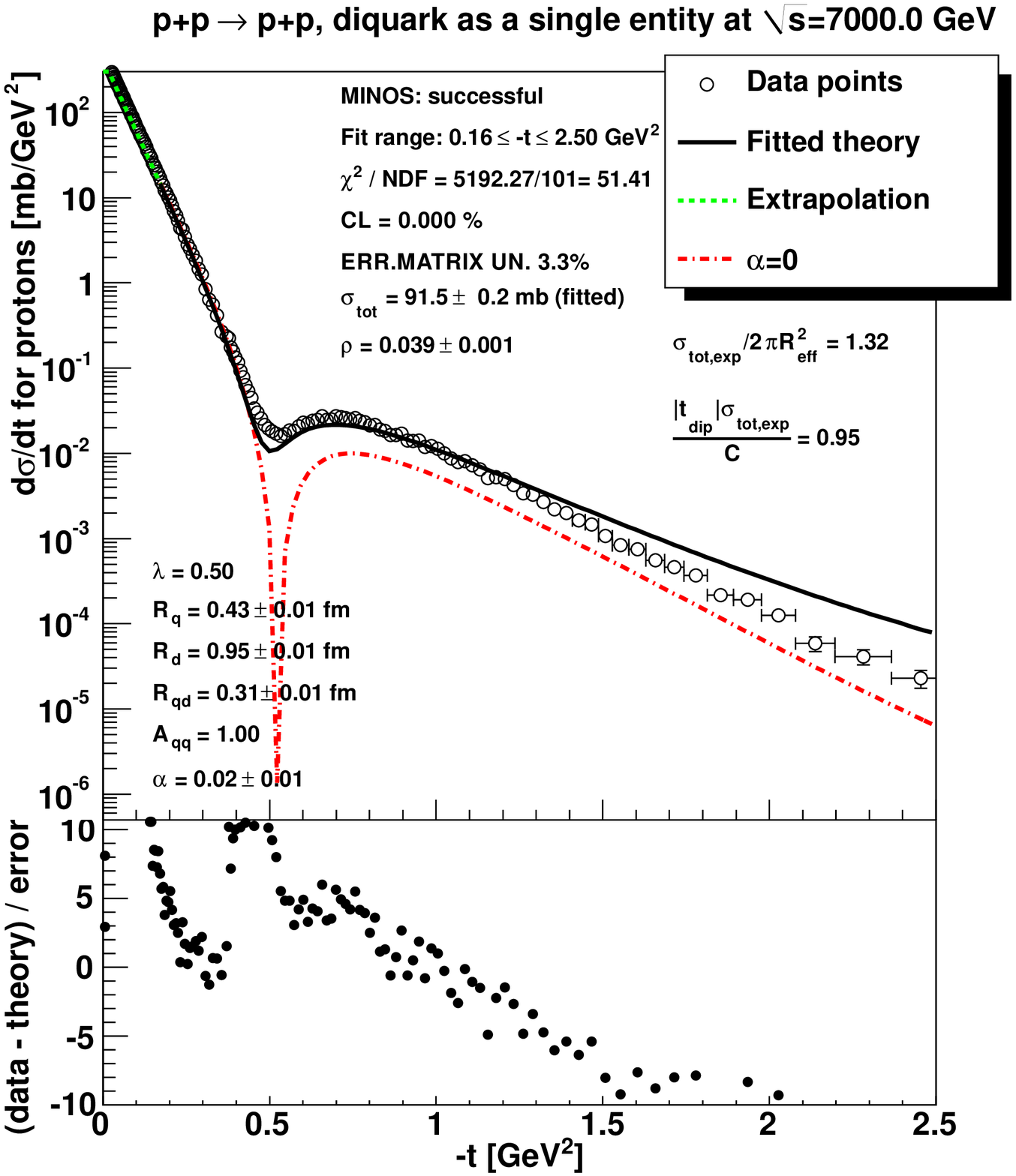}}
        \centering
        \caption{Fit result at $7$ TeV in the $0.16 \leq -t \leq 2.5$ GeV$^2$ range shown on the left hand side plot~(a). The case when the measured total cross-section is added to the fit as one additional data point
	is shown on the right hand side plot~(b). The diquark is assumed to be a single entity.}
        \label{overall_description_with_sigtot}
\end{figure}

Firstly, the $\alpha$BB model is compared to the interesting model of Grichine, Starkov and Zotov~\cite{Grichine:2012ry}, which is based on a quark-diquark representation of the
proton. In their model the amplitude of the proton-proton elastic scattering is described with one- and two-pomeron exchange between the quark and diquark constituents. In total
they apply only two fit parameters, which are the quark and diquark radii. However, due to the Pomeron parametrization, their theoretical curve misses the description of the
diffractive minimum. On Fig.~\ref{without_sigtot} our result is provided at $7$ TeV in the $0.16 \leq -t \leq 2.5$ GeV$^2$ range. This Fig.~\ref{without_sigtot}, as well as our Figs.~\ref{qq23}-\ref{qq7000} indicate,
that we  have improved significantly on the description of the dip region using a generalized Bialas-Bzdak model. Although the fit quality improved and become reasonable in the dip
region, we still could not find a reasonable description of the TOTEM dataset, that would work both in the low-$|t|$, in the dip and in the large-$|t|$ region. For example, if the low-$|t|$ region is
not included in the fit, this region is under-estimated by the extrapolated curve and also the total cross-section is under-estimated by $17.7$ \%. Fig~\ref{with_sigtot} indicates,
that if we include the measured total cross-section to the fit as an additional data point, but still keeping the fitted $t$ range to $0.16\le|t|\le2.5$ GeV$^{2}$, the
description improves at the low values of $|t|$ but the fit deviates more from the data in the dip region. The radius parameter of the quarks $R_q$ and that of the diquarks  $R_d$
together with the quark-diquark distance $R_{qd}$ increase  after adding the measured total cross-section.

The errors and the confidence levels are not provided in the paper of Grichine, Starkov and
Zotov~\cite{Grichine:2012ry}, however on a qualitative level apparently the $\alpha$BB model compares well with the inspiring  result of Grichine and collaborators.

\begin{figure}[H]
	\centering
	\includegraphics[trim = 0mm 4mm 0mm 4mm, clip, height=0.48\linewidth,width=0.48\linewidth]{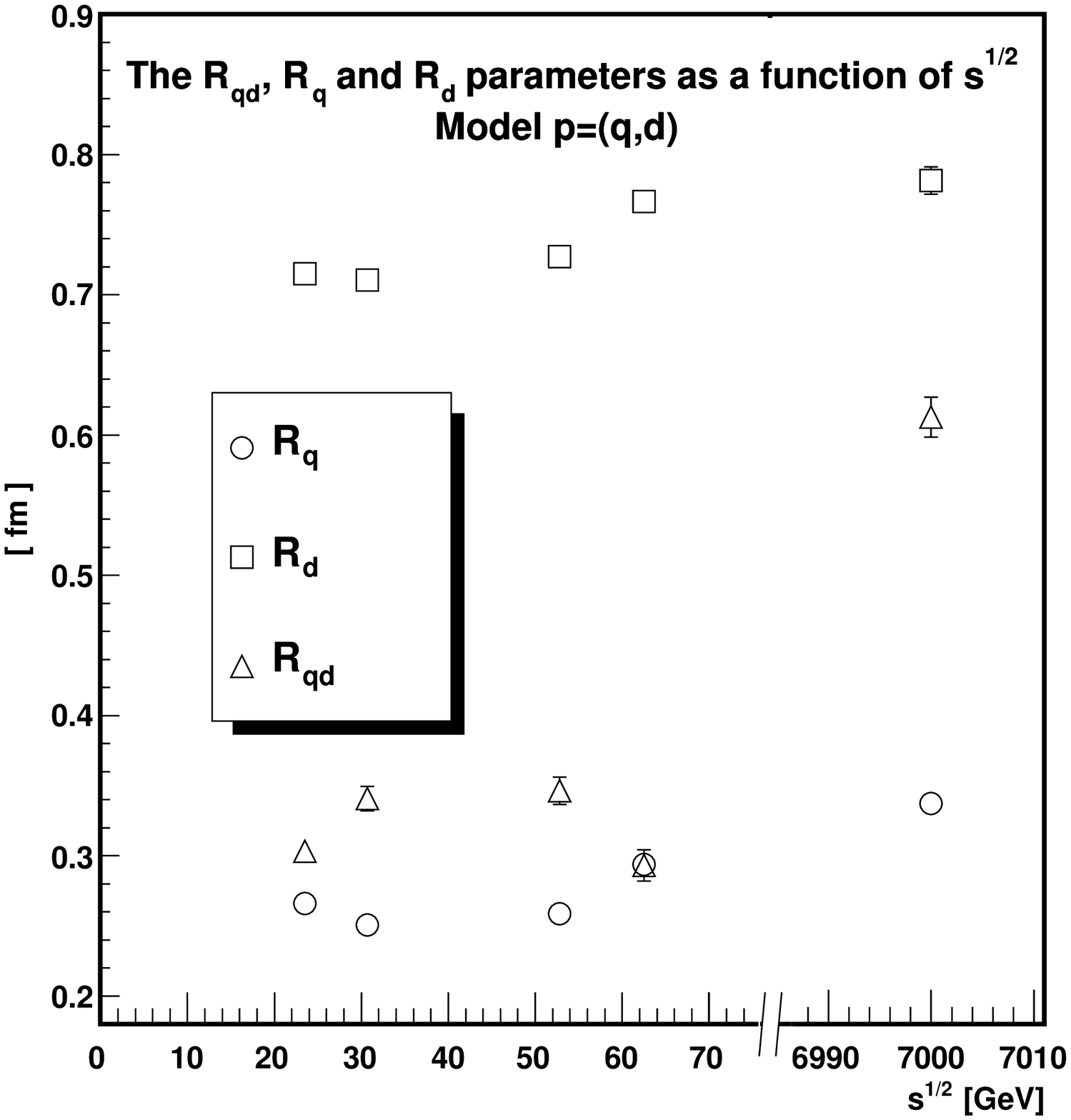}
	\includegraphics[trim = 0mm 4mm 0mm 4mm, clip, height=0.48\linewidth,width=0.48\linewidth]{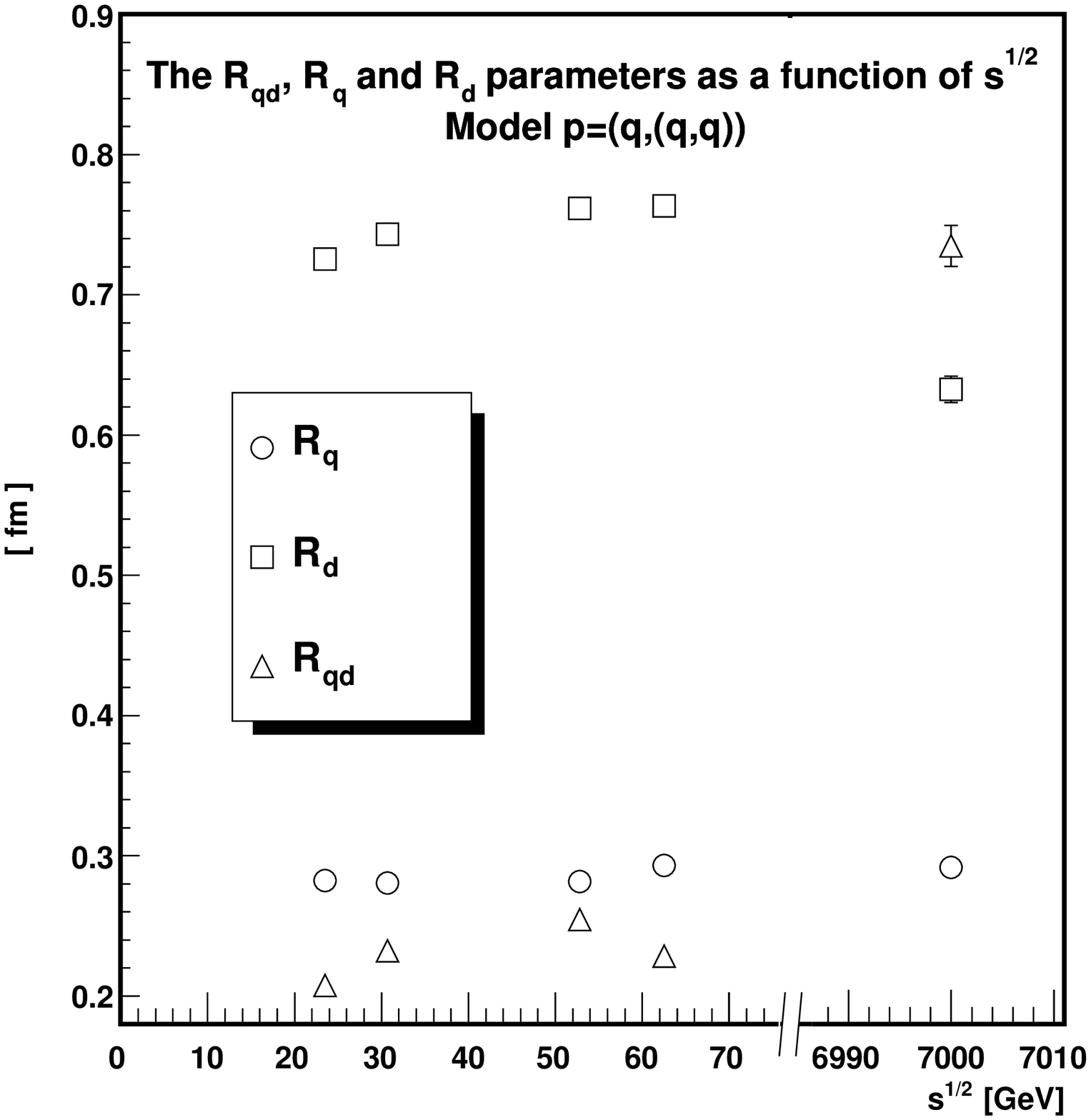}
	\caption{The $\sqrt{s}$ dependence of the parameters $R_{q}$, $R_{d}$, $R_{qd}$. The symbol size is greater than 
	the errors from the fit on several data points. Note that the most significant change is the increase of $R_{qd}$ at LHC energies.}
	\label{single_RqRdRqd}
\end{figure}

After the model comparison we turn to discuss the center of mass energy $\sqrt{s}$ dependence of our model parameters. The $\sqrt{s}$ dependence of the quark and diquark radius parameters $R_{q}$, $R_{d}$, together with the quark-diquark distance $R_{qd}$ is
shown on Fig.~\ref{single_RqRdRqd}, for
both the single and composite diquark model. The
results indicate that the quark and diquark radius parameters $R_{q}$ and $R_{d}$ are nearly constant, while the distance of the constituents $R_{qd}$ is increasing with increasing
collision energy $\sqrt{s}$.
\begin{figure}[H]
	\centering
        \subfigure{\includegraphics[trim = 0mm 4mm 0mm 8mm, clip, height=0.48\linewidth,width=0.48\linewidth]{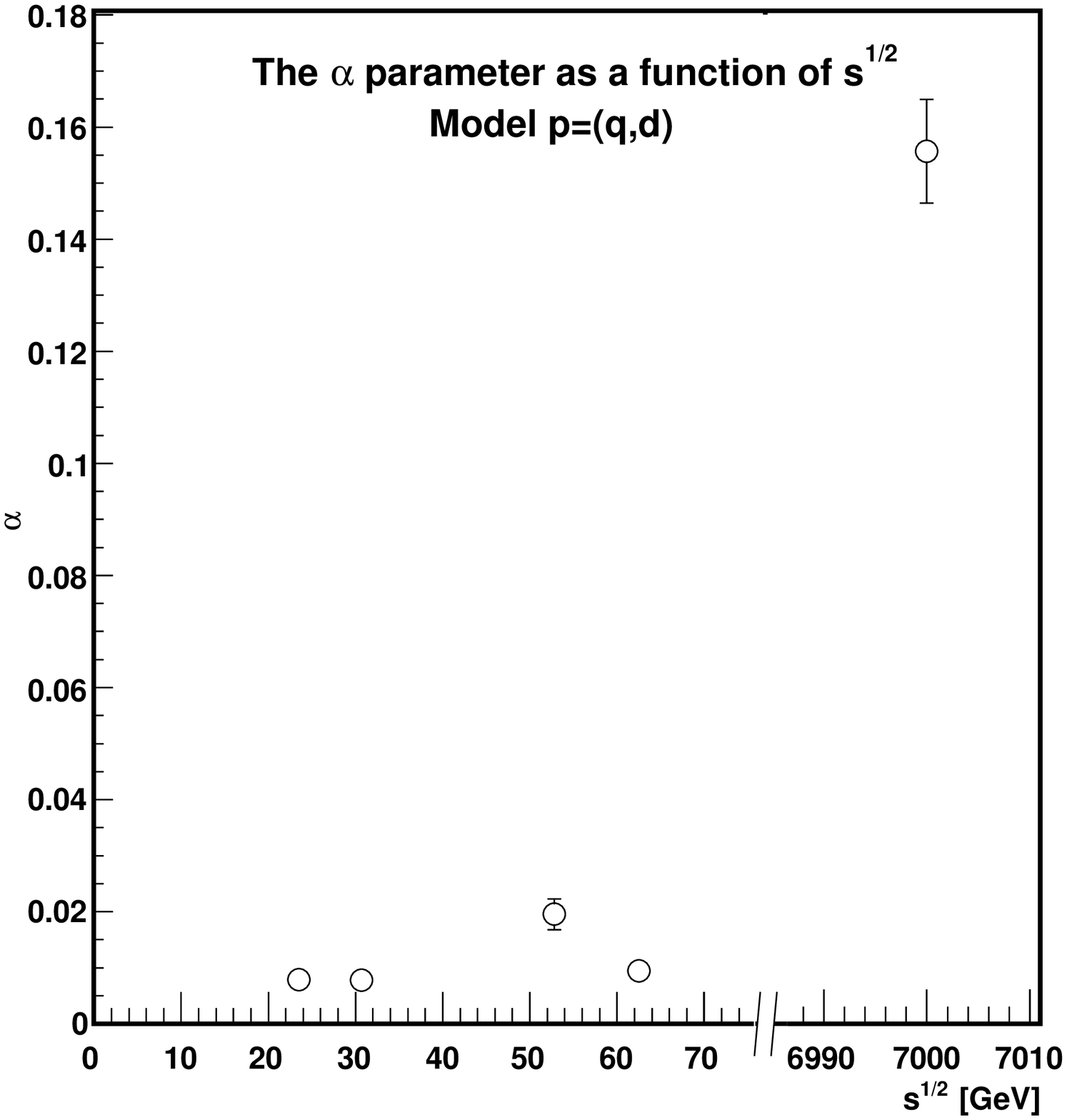}}
        \subfigure{\includegraphics[trim = 0mm 4mm 0mm 8mm, clip, height=0.48\linewidth,width=0.48\linewidth]{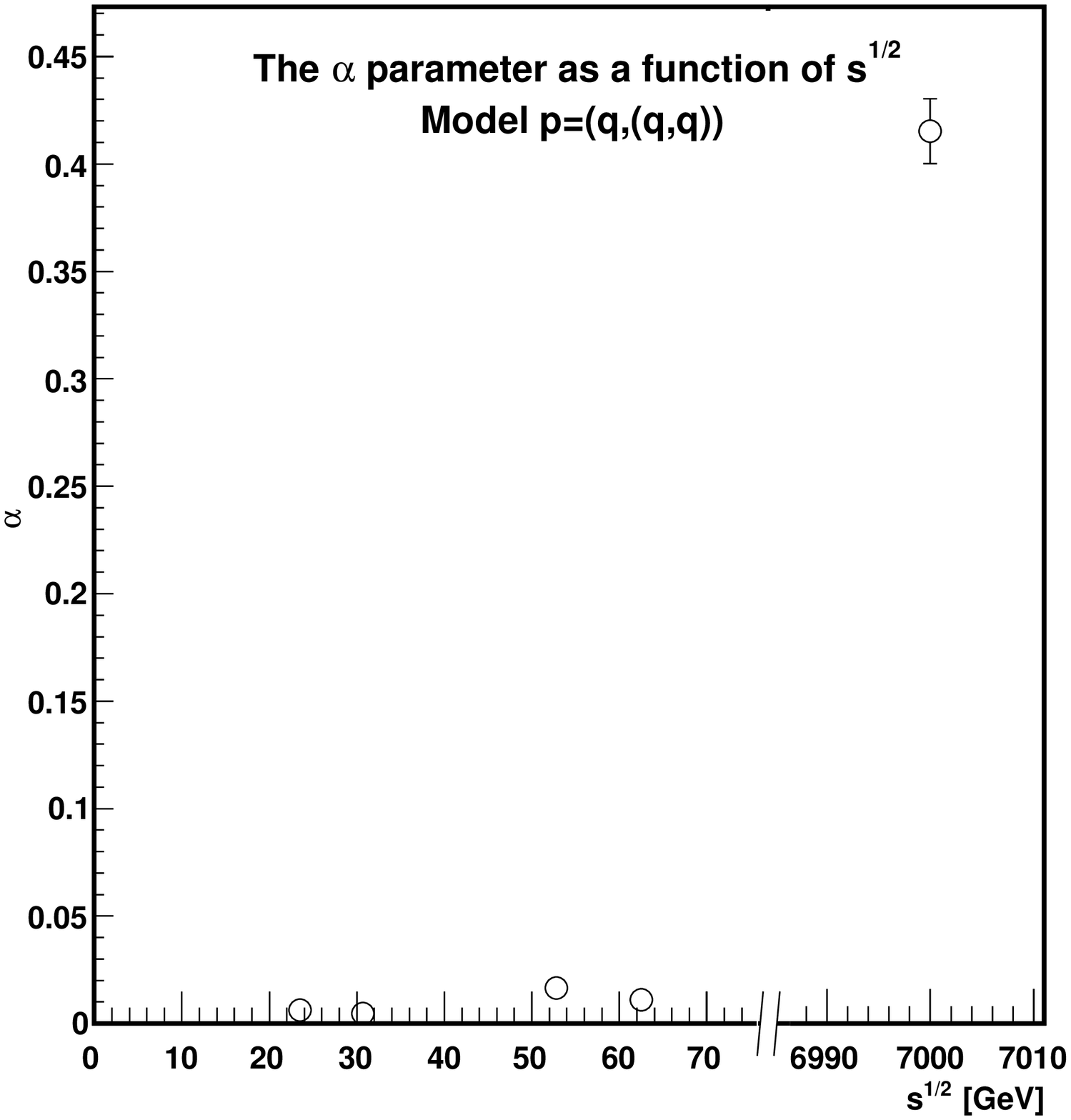}}
	\caption{The $\sqrt{s}$ dependence of the parameter $\alpha$. This plot indicates, that $\alpha$ becomes more important
	at LHC energies, than at ISR, indicating that it is more probable at LHC that a proton does not scatter elastically even if all of its constituents happened to scatter
	elastically. According to these results, the pp collisions become more and more ``fragile'' with increasing $\sqrt{s}$.}
	\label{alphaenergy}
\end{figure}

Although parameter $\alpha$ is introduced to have a successful data description at
the dip region, and it can be considered as the $\rho$ parameter of parton-parton level scattering (where $\rho$ is the ratio of the real to imaginary part
of the forward scattering amplitude at zero momentum transfer), our method to introduce $\alpha$ is based on an analogy with the Glauber-Velasco model and can be considered
valid in the leading order, $|\alpha|<<1$ limit only. Future work is needed to investigate, if terms that are non-linear functions of $\alpha$, can be derived that can improve
the agreement with the Bialas-Bzdak model not only in the dip but also in the whole experimentally available $|t|$ region at LHC energies.

\begin{figure}[H]
	\centering
        \subfigure{\includegraphics[height=0.48\linewidth,width=0.48\linewidth]{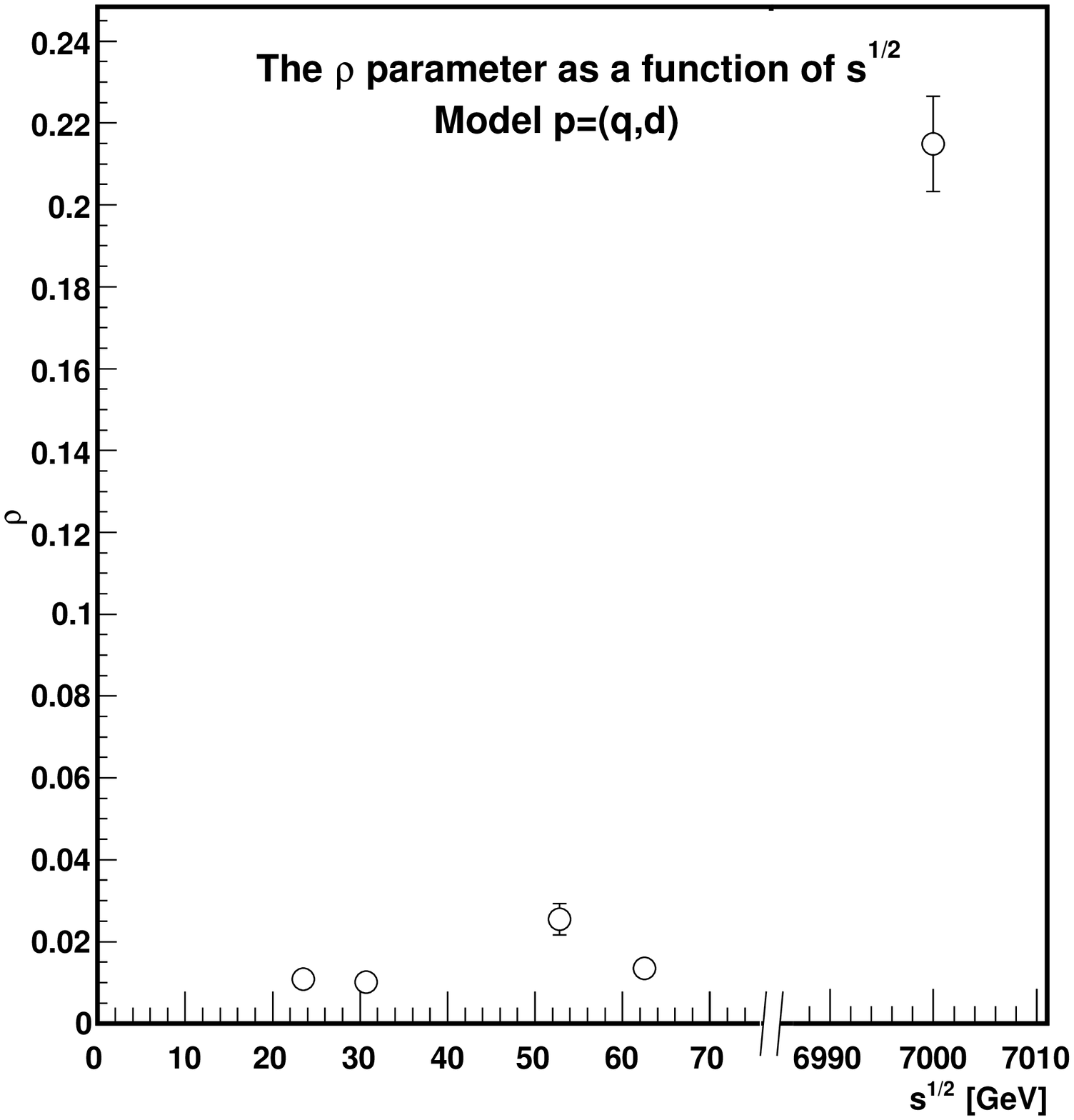}}
        \subfigure{\includegraphics[height=0.48\linewidth,width=0.48\linewidth]{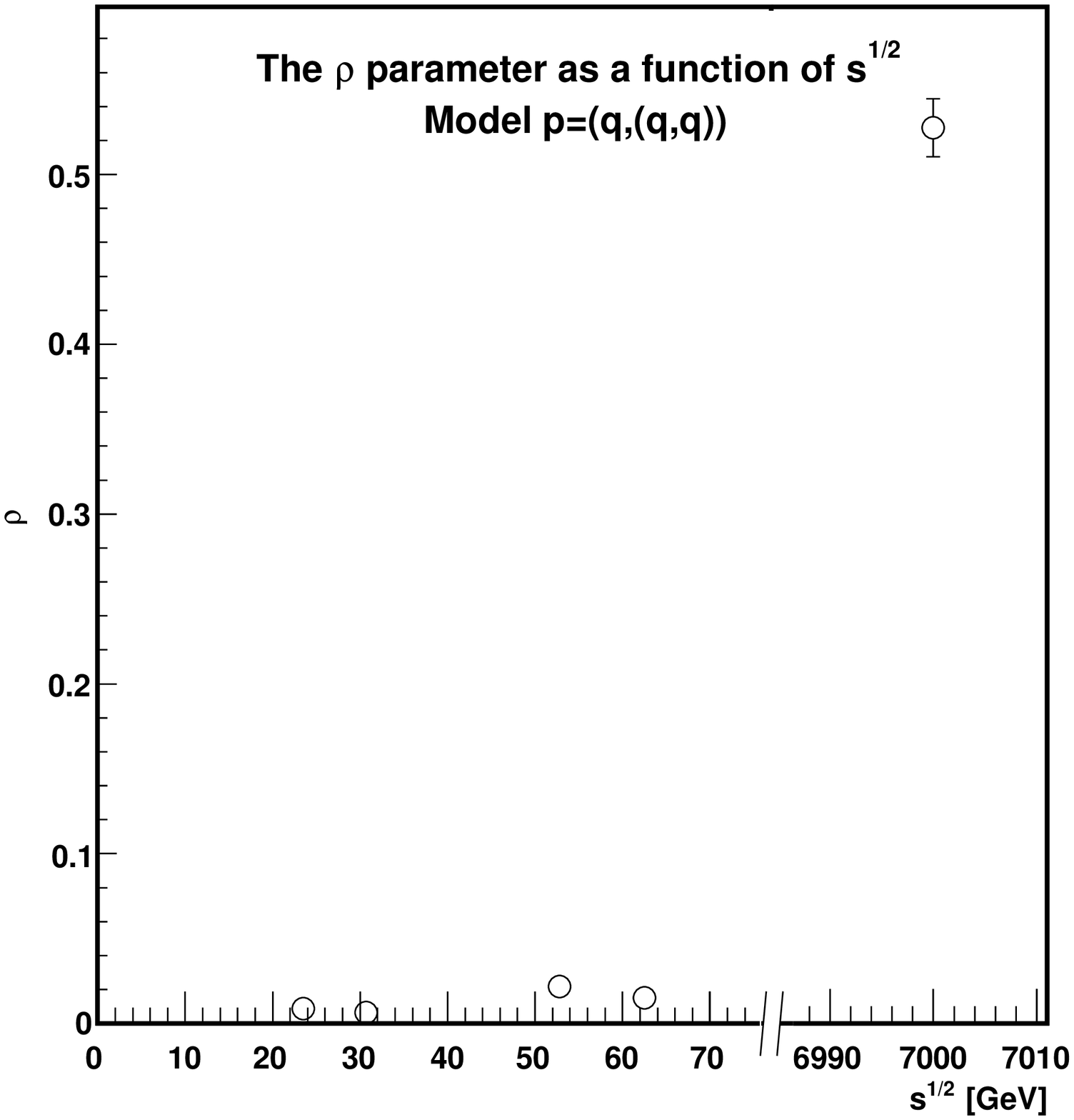}}

	\caption{The $\sqrt{s}$ dependence of the parameter $\rho$.}
	\label{rhoenergy}
\end{figure}

\begin{figure}[H]
        \centering
        \subfigure{\includegraphics[height=0.48\linewidth,width=0.48\linewidth]{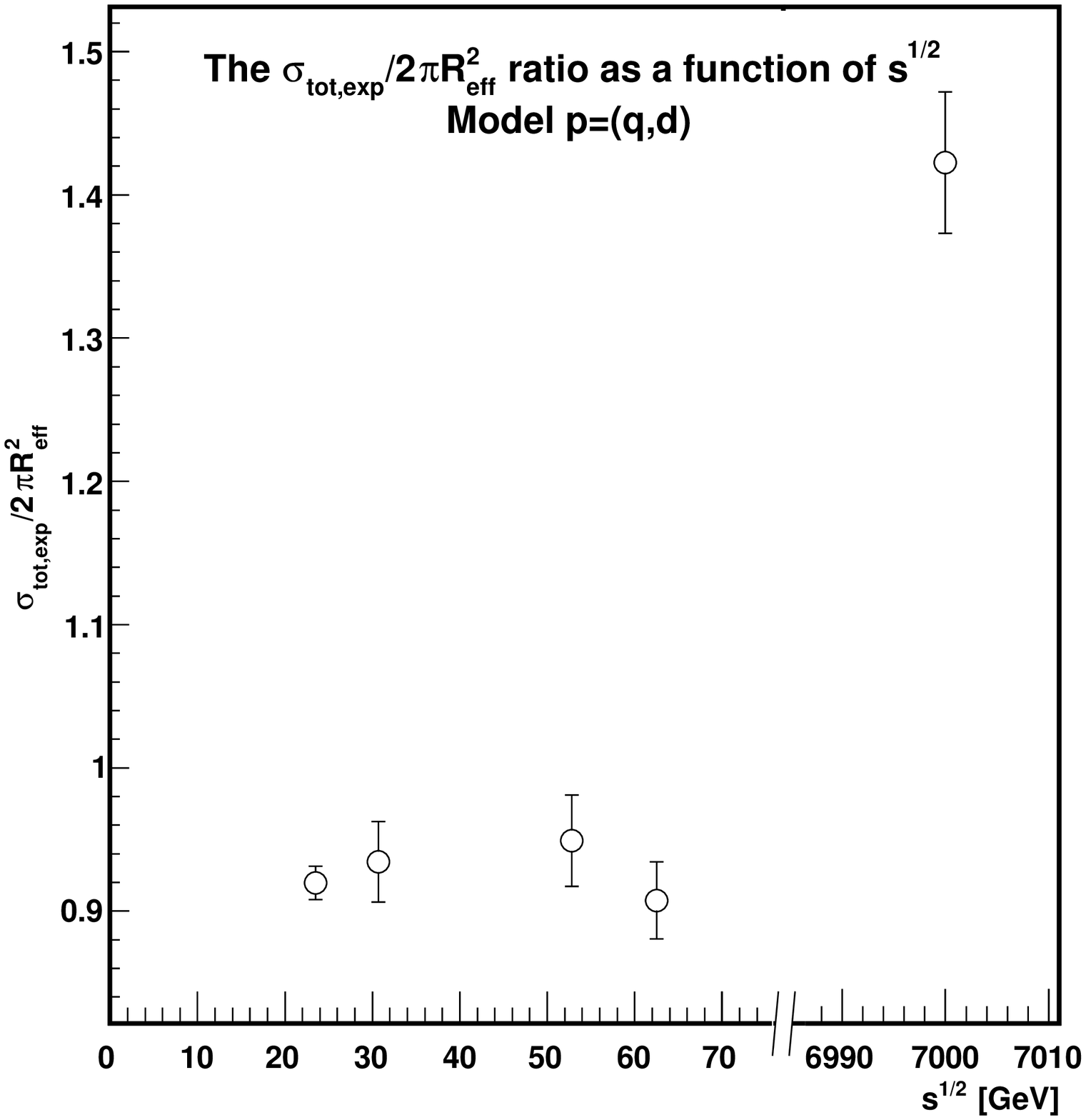}}
        \subfigure{\includegraphics[height=0.48\linewidth,width=0.48\linewidth]{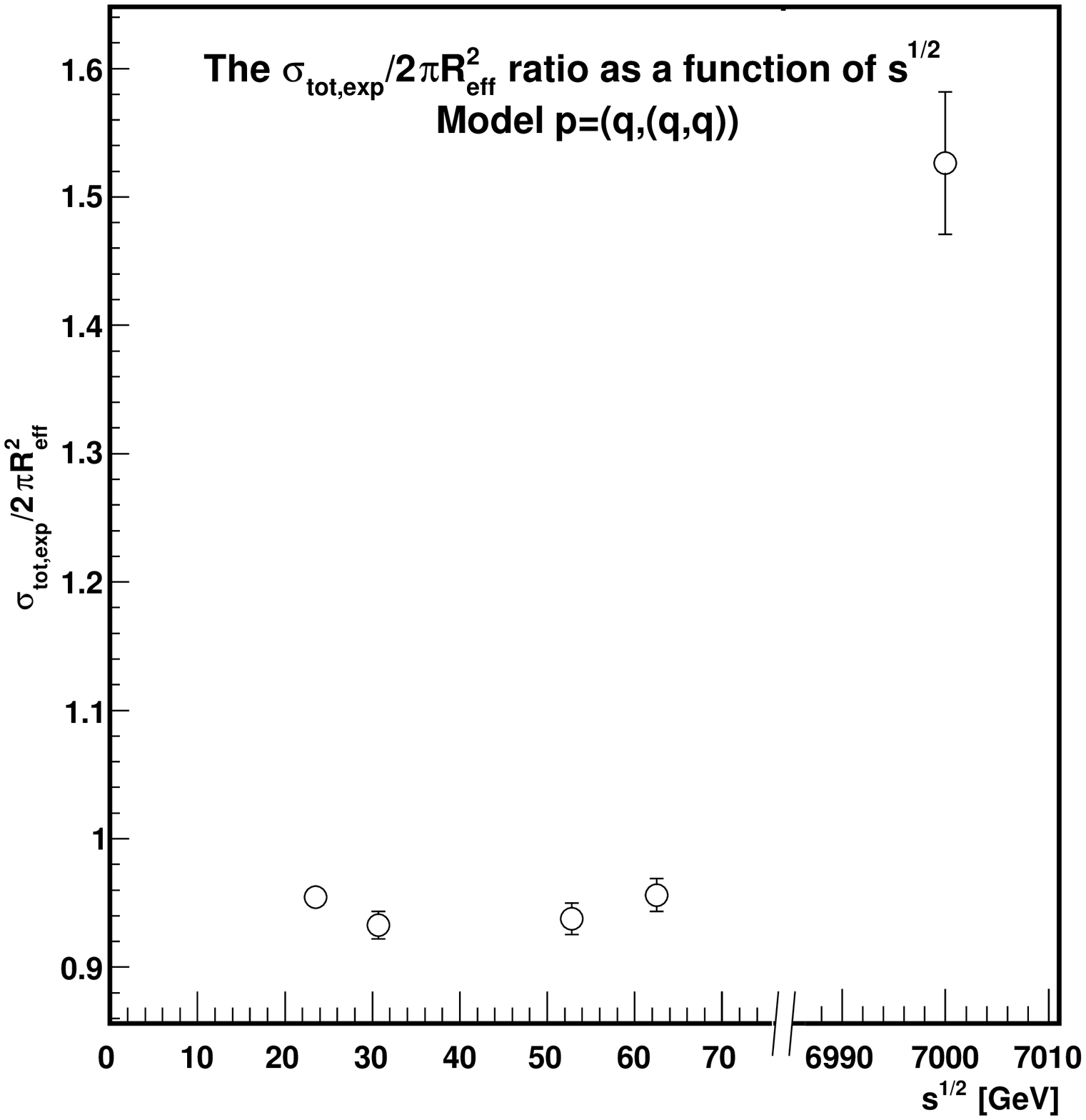}}
	\caption{The $\sigma_{tot,exp}/2\pi R_{eff}^{2}$ ratio.}
	\label{Reff_sigma_tot_ratio}
\end{figure}

The $\alpha$ parameter as a function of $\sqrt{s}$ is plotted on Fig. \ref{alphaenergy}, which shows the increasing role of this parameter at LHC energies. The
increase of $\alpha$ can be interpreted as the proton become more and more fragile with increasing $\sqrt{s}$. The evolution of the $\rho$ parameter is summarized on Fig.~\ref{rhoenergy} and
the obtained values are collected in Table~\ref{single_rho_phenom} and \ref{qq_rho_phenom}, for the single and composite diquark case, respectively. These two tables also contain the results of the
phenomenological relation Eq.~(\ref{eqReffpersigmatot}), which shows that the relation between the measured total cross-section $\sigma_{tot,exp}$ and the effective radius $R_{\text{eff}}$ is well satisfied
in case of ISR energies. Fig.~\ref{Reff_sigma_tot_ratio} indicates, that the $\sigma_{tot}/2\pi R_{\text{eff}}^{2}$ ratio is approximately constant, that is independent of $\sqrt{s}$ as well as from the particular
choice of the $p = (q,d)$  or the $p = (q, (q,q))$ model of the proton. The agreement is less convincing at $\sqrt{s}=7$ TeV, however, at this energy, the vanishing confidence level indicates, that the fit quality is not statistically acceptable. On
the other hand this fit still can be considered reasonable, and compares well to recent interesting results of Grichine and collaborators~\cite{Grichine:2012ry}. 

\begin{figure}[H]
	\centering
	\subfigure{\includegraphics[height=0.60\linewidth,width=0.55\linewidth]{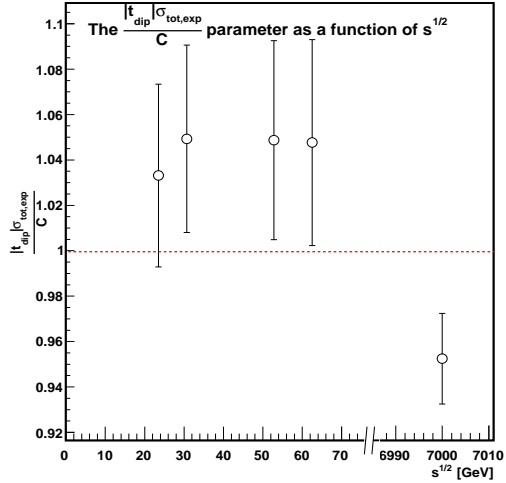}}
	\caption{The $\frac{|t_{dip}| \cdot \sigma_{tot,exp}}{C}$ ratio, directly obtained from experimental data. The dashed line indicates
	the ratio $1$.}
	\label{t_dip_sig_tot_figure}
\end{figure}

The second phenomenological relation Eq.~(\ref{t_dip_sig_tot}), which relates the position of the first diffractive minimum $t_{dip}$ and the total cross-section, is model independent, and well satisfied for
both ISR and LHC energies, according to Table~\ref{t_dip_sig_tot_table} and Fig.~\ref{t_dip_sig_tot_figure}. This relation~(\ref{t_dip_sig_tot}) was motivated by the formulae of photon scattering on a black disk,
where the elastic differential cross-section is~\cite{Block:2006hy}
\begin{equation}
	\frac{d\sigma_{black}}{dt}=\pi R^{4}\left[\frac{J_{1}(qR)}{qR}\right]^{2}\,,
	\label{dssigmadt_black}
\end{equation}
and the total cross-section
\begin{equation}
	\sigma_{tot,black}=2\pi R^{2}\,.
	\label{sigmatot_black}
\end{equation}
In this simple theoretical model the position of the first diffractive minimum, following from Eq.~(\ref{dssigmadt_black}), and the total cross-section Eq.~(\ref{sigmatot_black}) satisfies 
	\begin{equation}
		C_{black}=|t_{dip,black}|\cdot\sigma_{tot,black}=2\pi j_{1,1}^{2}(\hbar c)^2\approx~35.9\,\text{mb GeV}^{2}\,,
		\label{Cblack}	 
	\end{equation}
where $j_{1,1}$ is the first root of the Bessel function~$J_{1}(x)$.

In case of the phenomenological relation  Eq.~(\ref{t_dip_sig_tot}) the constant $C$ was fitted to the measured data, to obtain the best possible description. It is clear that this fitted constant $C=54.8\pm0.7\,$mb GeV$^{2}$ is significantly different from the
number $C_{black}$ of Eq.~(\ref{Cblack}) one may expect from light scattering on a black disc.

In this sense, although Eq.~(\ref{t_dip_sig_tot})
is satisfied by ISR as well as TOTEM data, the value of the constant indicates a more complex scattering phenomena, than the photon black disc scattering, mentioned above.

The observed $|t_{dip}|\,\sigma_{tot} \approx C$ relationship may in fact be a reflection of a deeper scaling property of the differential cross section
of elastic p+p scattering. To guide our intuition, it is again useful to consider the scattering of light on a black disk. The differential cross-section
(\ref{dssigmadt_black}) can be expressed in terms of the momentum transfer $t=-q^2$
\begin{equation}
	\frac{d\sigma_{black}}{dt}=\frac{\pi R^{2}}{|t|}J_{1}(\sqrt{|t|}R)^{2}=\frac{\sigma_{tot,black}}{2|t|}J_{1}\left(\sqrt{\frac{|t|\sigma_{tot,black}}{2\pi}}\right)^{2}\,.
	\label{dssigmadt_black_2}
\end{equation}
The result Eq.~(\ref{dssigmadt_black_2}) can be scaled to a universal scaling curve
\begin{equation}
	\frac{|t|}{\sigma_{tot,black}} \frac{d\sigma_{black}}{dt} = \frac{1}{2} J_{1}\left(\sqrt{\frac{|t| \sigma_{tot,black}}{2\pi}}\right)^{2}=\frac{1}{2} J_{1}\left(\sqrt{\frac{y}{2\pi}}\right)^{2}=F_{black}(y),
	\label{Fy}
\end{equation}
where $F_{black}(y)$ is a dimensionless function of the variable $y=|t|\sigma_{tot,black}$, and is the same for all black discs regardless of their radius. The
function $F_{black}(y)$ can be expressed in terms of the dimensionless variable $z=|t|/|t_{dip,black}|=y/C_{black}$. From Eq.~(\ref{Cblack}) and Eq.~(\ref{Fy})
\begin{equation}
	\frac{|t||t_{dip,black}|}{C_{black}} \frac{d\sigma_{black}}{dt} = \frac{1}{2} J_{1}\left(\sqrt{\frac{zC_{black}}{2\pi}}\right)^{2}=G_{black}(z)\,,
	\label{Gz}
\end{equation}

\begin{figure}[H]
        \includegraphics[width=0.49\linewidth]{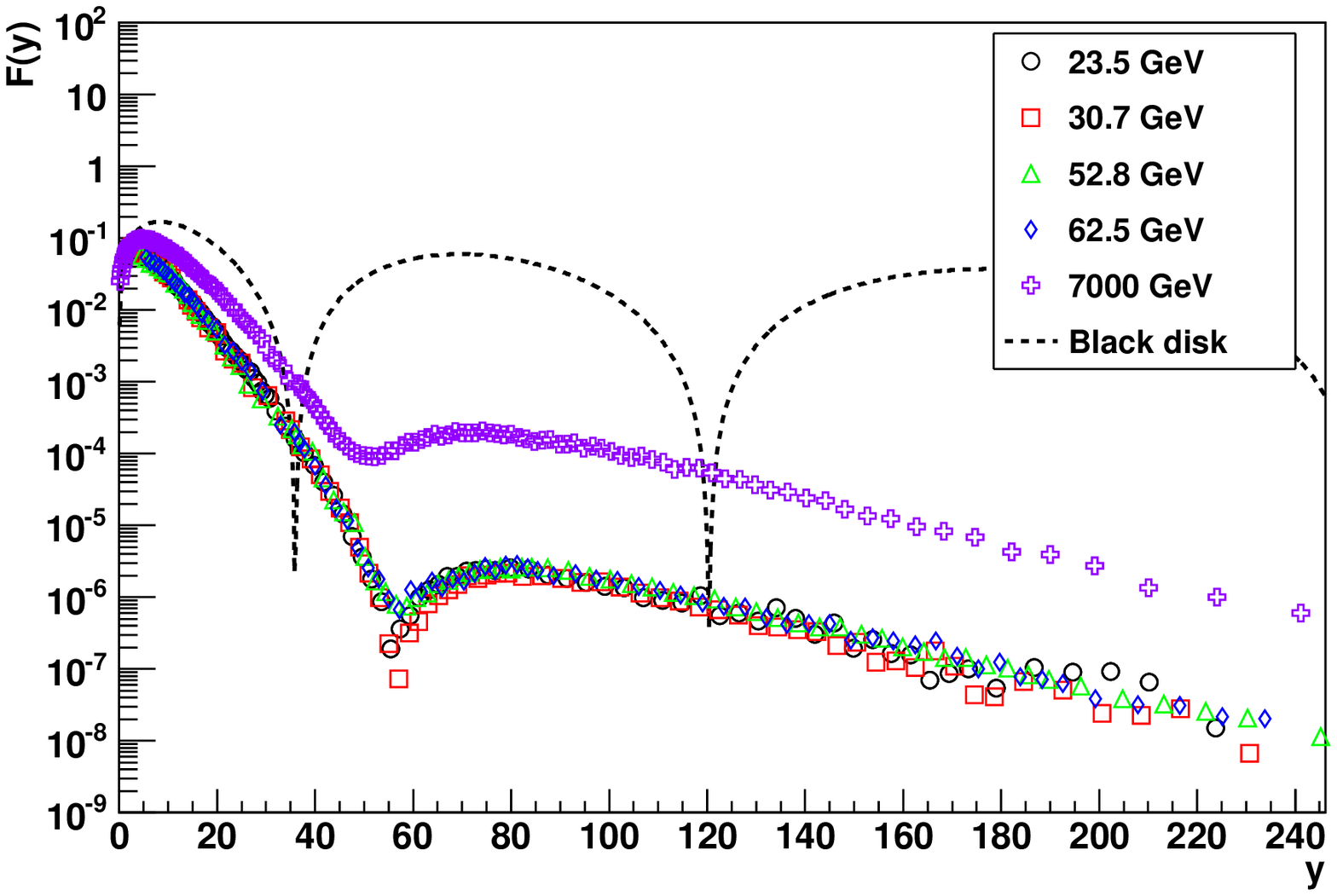}
	\includegraphics[width=0.49\linewidth]{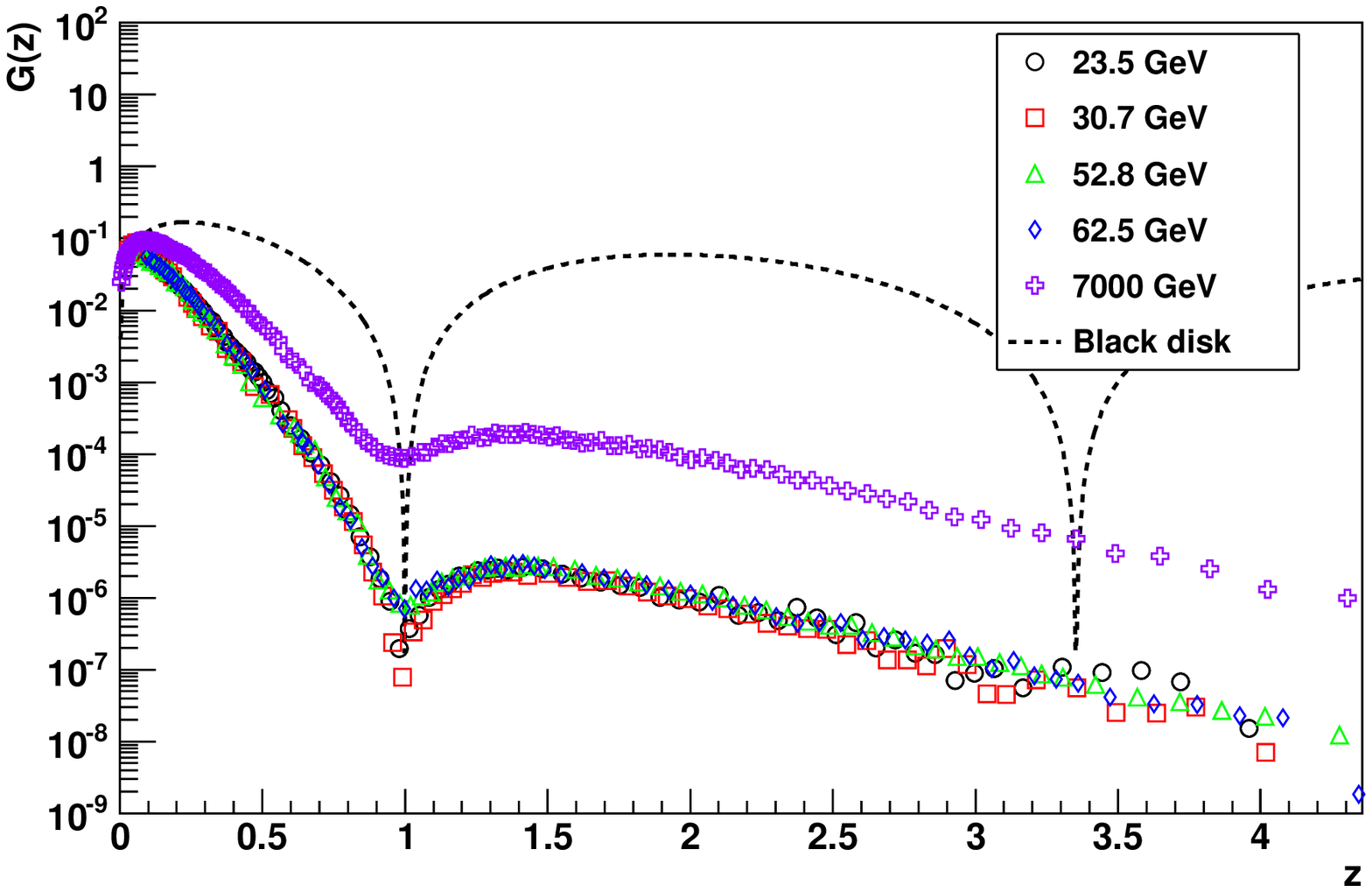}
        \centering
        \caption{The $F(y)$ and $G(z)$ scaling functions showing the ISR and TOTEM data. The scaling functions of the black disc $F_{black}(y)$ and $G_{black}(z)$ are also shown.}
        \label{Fy_and_Gz}
\end{figure}

Both of the above dimensionless functions can be generalized to the real experimental case, leading to $F(y)$ and $G(z)$, where $y=|t|\sigma_{tot}$ and $z=y/C$.  We have thus plotted both the $F(y)$ and
the $G(z)$ functions for all the ISR and LHC data available for us. The results are shown on Fig.~\ref{Fy_and_Gz}, where the scaling functions of the black disc $F_{black}(y)$ and $G_{black}(z)$ are also presented. The
plots indicate, that ISR and LHC data on $d\sigma/dt$  approximately satisfy these newly found scaling relation, but with some scaling violating terms.

The data indicate that the proton-proton elastic differential cross-section $d\sigma/dt$ data collapse to an energy independent scaling function at the ISR energies of 23.5 - 62.5 GeV, and data
at $\sqrt{s} = 7$ TeV are significantly and qualitatively
different, move closer to the scaling functions that characterize the black disc limit. However, even at $\sqrt{s} = 7$ TeV, the data are significantly
different from the scaling function of the black disc limit. In particular, the
secondary minimum of the black disc limit is not observed, so scaling violating terms are still important. It remains to be seen if this trend of approaching better the black
disc limit continues with increasing colliding energies, or not. Plotting the $F(z)$ and the $G(z)$ scaling functions seems to be a useful tool to investigate, to understand 
how the data approach this possible limiting behaviour.

The purpose of our manuscript was to improve  the Bialas-Bzdak model in the dip region,
where the original model was singular. Indeed, we report on a successful and significant improvement
of the data description in the dip region from a generalized Bialas-Bzdak model in this manuscript. We do
not claim several things: we for example do not claim that this is the only imaginable model and possible
improved other models, like models with $n$ Pomerons and $m$ Odderons should not be considered in the future.
As there are several possibilities to improve on the description of the TOTEM data, our choice to improve on the Bialas-Bzdak model
by adding a small real part to the forward scattering amplitude is just one of the possibilities even if this consideration is based on a physical motivation.

\begin{table}[!ht]
\centering
\begin{tabular}{|c|c|c|c|c|c|c|c|c|c|} \hline
$\sqrt{s}$ [GeV] & 23.5				& 30.7				& 52.8				& 62.5				& 7000					\\ \hline\hline
$\rho$ &0.01 		 $\pm$ 0.01 	& 0.01 		 $\pm$ 0.01 	& 0.03 		 $\pm$ 0.01 	& 0.01 		 $\pm$ 0.01 	& 0.21 		 $\pm$ 0.01	 	 \\ \hline
$\sigma_{tot,exp}/2\pi R_{eff}^{2}$ &0.92 		 $\pm$ 0.01	& 0.93 		 $\pm$ 0.03	& 0.95 		 $\pm$ 0.03	& 0.91 		 $\pm$ 0.03	& 1.42 		 $\pm$ 0.05	 	 \\ \hline
\end{tabular}
\caption{The $\rho$ parameter and the $\sigma_{tot,exp}/2\pi R_{eff}^{2}$ ratio in case the diquark is assumed to be a single entity. Fit range is the same $0.36\leq -t \leq 2.5$ GeV$^2$ for all datasets.
    Note that the fit quality is not acceptable at $\sqrt{s}=7$ TeV.}
    \label{single_rho_phenom}
\end{table}

\begin{table}[!ht]
\centering
\begin{tabular}{|c|c|c|c|c|c|c|c|c|c|} \hline
$\sqrt{s}$ [GeV] & 23.5				& 30.7				& 52.8				& 62.5				& 7000					\\ \hline\hline
$\rho$ &0.01 		 $\pm$ 0.01 	& 0.01 		 $\pm$ 0.01 	& 0.02 		 $\pm$ 0.01 	& 0.02 		 $\pm$ 0.01 	& 0.53 		 $\pm$ 0.02	 	 \\ \hline
$\sigma_{tot,exp}/2\pi R_{eff}^{2}$ &0.95 		 $\pm$ 0.01 	& 0.93 		 $\pm$ 0.01	& 0.94 		 $\pm$ 0.01	& 0.96 		 $\pm$ 0.01	& 1.53 		 $\pm$ 0.06	 	 \\ \hline
\end{tabular}
\caption{The $\rho$ parameter and the $\sigma_{tot,exp}/2\pi R_{eff}^{2}$ ratio, the diquark is assumed to be a composite entity.  Fit range is the same $0.36\leq -t \leq 2.5$ GeV$^2$ for all datasets.
Note that the fit quality is not acceptable at $\sqrt{s}=7$ TeV.}
    \label{qq_rho_phenom}
\end{table}

\begin{table}[!ht]
\centering
\begin{tabular}{|c|c|c|c|c|c|c|c|c|c|} \hline
$\sqrt{s}$ [GeV] & 23.5				& 30.7				& 52.8				& 62.5				& 7000					\\ \hline\hline
$\frac{|t_{dip}|\,\sigma_{tot,exp}}{C}$ &1.03              $\pm$ 0.04     & 1.05           $\pm$ 0.04     & 1.05           $\pm$ 0.04     & 1.05           $\pm$ 0.05     & 0.95           $\pm$ 0.02	 	 \\ \hline
\end{tabular}
\caption{The $\frac{|t_{dip}|\,\sigma_{tot,exp}}{C}$ ratio,
        model independently. Precise up to $5\,\%$ (within errors).}
	\label{t_dip_sig_tot_table}
\end{table}

One can ask the question, why and at what range of four momentum transfer $|t|$ such a simple model should work. Apparently, for sufficiently large $|t|$ we start to probe a distance that becomes smaller than $0.2$ fm, the size of a constituent quark
in the Bialas-Bzdak picture, or, the even larger diquark scale.  Thus, for a large enough $|t|$ the simple Bialas-Bzdak model, and even its generalized version presented in the current manuscript, is expected to fail. Indeed, TOTEM observed that above
the dip structure the functional form of the differential cross-section changes, and can be described with a power law $|t|^{-n}$ with an exponent $n=7.8\pm0.3^{\text{stat}}\pm0.1^{\text{syst}}$ for $|t|$-values between 1.5 GeV$^{2}$ and 2.0 GeV$^{2}$~\cite{Antchev:2011zz}.

This TOTEM result on the large $|t|$ elastic scattering is consistent with the perturbative QCD prediction, based on a spin-1 gluon exchange picture, that predicted an energy independent power-low
tail of the $d\sigma/dt$ distribution with $n=8.0$~\cite{Donnachie:1979yu}.

\section{Conclusion and outlook}
In this work we have generalized the geometrical Bialas-Bzdak model of elastic proton-proton scattering by allowing for a real part
of the forward scattering amplitude using the same geometrical picture, but assuming that a proton-proton scattering may become
inelastic even in the case when all scattering of the constituents of the colliding protons is elastic (but not completely collinear).
This generalization resulted in a successful description of the dip region of elastic proton-proton scattering in the ISR energy region
and resulted in a significant, qualitative improvement of the ability of this model to describe elastic proton-proton scattering at
7 TeV colliding energies as measured by the TOTEM Collaboration at CERN LHC. We have also found that the generalized Bialas-Bzdak 
model can describe only the small $|t|$ data set and therefore the total cross-section $\sigma_{tot}$ at LHC, if the fit range is
limited to a relatively low $|t|$ range. However,  even this generalized Bialas-Bzdak model fails,
if the large $|t|$ region data of TOTEM is included, more precisely we find no good quality fits in the
$0.0 \leq -t \leq a$ GeV$^2$ region, where $a>0.8$ GeV. This result can be interpreted as a qualitative change in elastic 
proton-proton scattering at the 7 TeV LHC energies as compared to the top ISR energies, due to the opening of a new channel in this
reaction. It would be interesting to collect data at RHIC and at lower LHC energies to determine the energy region where 
this transition happens and the new physics channel opens.

Based on the geometrical picture behind the Bialas-Bzdak model, we have identified
and tested the validity of two simple phenomenological formulas. Our first formula relates the total proton-proton scattering
cross-section to an effective radius, that is the quadratic sum of the radii of a quark and a diquark as well as the distance between the 
center of mass of the diquark and the quark inside the proton. Regardless of the detailed structure of the diquarks, and independently of 
the values of the real part of the forward scattering amplitude, this formula gives a model independent estimate for the total
cross-section with a typical 10 \% precision at ISR energies, that becomes worse at LHC energies, but  our data analysis nevertheless
suggests that the increase of the total cross-section of proton-proton scattering is mainly due to the increase of the quark-diquark
separation with increasing colliding energies, while the size of the constituent quarks and diquarks is approximately independent
of the colliding energies. Our second formula establishes a relation between the total cross-section of proton-proton scattering and the
position of the dip in the differential cross-section: in particular, we find that the product of these two quantities is a model independent
constant. We also demonstrated that this formula is remarkably precise, it is satisfied by the ISR data within one
standard deviations, while at LHC the formula is also satisfied by the data within 3 standard deviations.

Given that the energy dependence of the total cross-section is very well described by linear or quadratic polynomials in $\ln({s})$,
or by Glauber Monte-Carlo simulations in case of proton-nucleus and nucleus-nucleus collisions, our result can in principle be well 
used to predict the position of the dip in the differential cross section of elastic scattering at various energies and colliding systems.
A detailed application of this relation to predict the dip position in p-Pb elastic scattering will be presented elsewhere~\cite{CsMCsTNFNT}.

\section{Acknowledgement}

T.\,Cs. would like to thank Prof.~R.\,J.\,Glauber inspiring discussions and for his kind hospitality at Harvard University. The authors
are also grateful to G.~Gustafson and M.~Csan\'ad for valuable discussions. The authors would like to thank also to M.~Giordano for valuable
comments, criticism and suggestions.

This research was supported by a Ch. Simonyi Fellowship as well as by the Hungarian OTKA grant NK 101438.

\end{document}